\documentclass[useAMS,usenatbib]{mn2e}
\usepackage{latexsym,amsmath,amssymb,enumerate}
\usepackage[]{graphicx,lscape}
\usepackage{subfigure}
\usepackage{multirow}

\usepackage{color,verbatim}
\usepackage{ulem}
\DeclareMathOperator{\bias}{Bias}
\DeclareMathOperator{\Var}{Var}

\title[Resolving the mass--anisotropy degeneracy II]{ 
Resolving the mass--anisotropy degeneracy of the spherically symmetric Jeans equation II: optimum smoothing and model validation}
\author[F. I. Diakogiannis, G. F. Lewis and R. A. Ibata]{
Foivos I. Diakogiannis$^{1}$\thanks{E-mail:
f.diakogiannis@physics.usyd.edu.au}, Geraint F. Lewis$^{1}$    and Rodrigo A. Ibata$^{2}$\\
$^{1}$Sydney
Institute for Astronomy, School of Physics, A28, University of Sydney, NSW 2006, Australia\\
$^{2}$Observatoire
Astronomique, Universit\'{e} de Strasbourg, CNRS, 11, rue de l Universit\'{e}, F-67000 Strasbourg, France}
\begin{document}

%\date{ {\bf CHANGE!!!} Accepted 1988 December 15. Received 1988 December 14; in original form 1988 October 11}

\pagerange{\pageref{firstpage}--\pageref{lastpage}} \pubyear{2013}

\maketitle

\label{firstpage}
\begin{abstract}
The spherical Jeans equation is widely used to estimate the mass content of a stellar systems with apparent  spherical symmetry. However, this method suffers from a degeneracy 
between the assumed mass density and the kinematic anisotropy profile, $\beta(r)$.  
In a previous work, we laid the theoretical foundations for an algorithm that combines smoothing B-splines with equations from dynamics to remove this degeneracy. Specifically, our method reconstructs a unique kinematic profile of $\sigma_{rr}^2$ and $\sigma_{tt}^2$ for an assumed free functional form  of the potential and mass density $(\Phi,\rho)$ and given a set of observed line-of-sight velocity dispersion measurements, $\sigma_{los}^2$.  
In Paper I (submitted to MNRAS: MN-14-0101-MJ) we demonstrated the efficiency of our algorithm with a very simple example  and we commented on the need for optimum smoothing of the B-spline representation; this is in order to avoid unphysical variational behaviour when we have large uncertainty in our data. 
In the current contribution we  present a process of finding the optimum smoothing for a given data set by using information of the behaviour from known ideal theoretical models. 
Markov Chain Monte Carlo methods are used to explore the degeneracy in the dynamical modelling process.
We validate our model through  applications to synthetic data for  systems with constant or variable mass-to-light ratio $\Upsilon$.
In all cases we recover excellent fits of theoretical functions to observables and unique solutions.
Our algorithm is a robust method for the  removal of the mass-anisotropy degeneracy of the spherically symmetric Jeans equation for an assumed functional form of the mass density. 
\end{abstract}


\section{Introduction}

An open problem in modern theoretical galactic dynamics is the mass anisotropy degeneracy of the spherically symmetric Jeans equation (hereafter SSJE). Specifically, there exists a degeneracy between an assumed mass density, $\rho(r)$, and the anisotropy profile, $\beta(r)$, of a self gravitating system. That is, there are many pairs $(\rho(r),\beta(r))$ that describe equally well the observables
\citep{1987ApJ...313..121M}. 
\cite{1982MNRAS.200..361B} were the first to present a solution for the case of constant mass-to-light ratio. 
From a different perspective, in a previous work (Diakogiannis et. al. 2014  submitted to MNRAS: MN-14-0101-MJ; hereafter Paper I), we set the theoretical foundation for the removal of this anisotropy so that: for an assumed free functional form\footnote{By the phrase ``assumed free functional form'' we mean a specific mass profile, e.g. King, or Michie, with free parameters to be estimated from our algorithm.} of the mass density and potential of a spherically symmetric system $(\rho(r), \Phi(r))$ we recover a unique kinematic profile for the second order radial, $\sigma_{rr}^2$, and tangential, 
$\sigma_{tt}^2$, velocity moments. This is valid for both constant and variable mass-to-light ratio.  
In Paper I  we argued that, if we know the complete functional form of $\sigma_{los}^2$ and the mass density $\rho(r)$, then it is possible by using some optimization method (e.g. Genetic Algorithms), to reconstruct a unique kinematic profile $\sigma_{rr}^2,\sigma_{tt}^2$  within any desired numerical accuracy; i.e. a unique decomposition of $\sigma_{los}^2$ to $\sigma_{rr}^2$ and $\sigma_{tt}^2$. 
For the case where we have a discrete data set, we can only speak of a unique family of profiles within some statistical uncertainty.  This uniqueness  follows from exploring parameter space with MCMC methods and recovering unimodal marginalized distributions. 

In the core of our method lies the representation of the radial velocity dispersion,  $\sigma_{rr}^2$,  in a B-spline basis. B-spline functions \citep{de1978practical} are used extensively in Computer Aided Geometric Design  
\citep{Farin:2001:CSC:501891} and in statistical smoothing spline modelling  techniques \citep{hastie01statisticallearning}.  
In Paper I we  demonstrated our algorithm with a very simple example with fixed  mass-to-light ratio $\Upsilon$. In the present paper we validate our method by giving a variety of examples with constant  or variable $\Upsilon$. For the latter we demonstrate the usage of our algorithm in a two component system  that consists of both stellar and dark matter populations. 
In Paper I we also commented on the need for optimum smoothing, since there may be cases where our data have large errors and the resulting kinematic profile demonstrates unphysical variations. In the current contribution we present an algorithm for optimum smoothing, based on information of the smoothness behaviour from ideal theoretical models.
 
The structure of our paper is as follows: 
in Section \ref{DLI_BSplines_MathForm} we present for completeness a very short description of the most basic equations presented in Paper I. 
In Section \ref{DLI_BSplines_StatAnal} we describe the statistical inference methods we use as well as the algorithm for optimised  smoothing. In Section \ref{DLI_BSplines_Examples} we present three detailed examples. In these we reconstruct fully the mass content of the system and the kinematic profile, using synthetic  data of brightness and line-of-sight velocity dispersion $\sigma_{los}^2$. In Section \ref{DLI_BSplines_Discussion} we discuss various aspects of our results. Finally in Section \ref{DLI_BSplines_Conclusions} we conclude our work.

\section{Mathematical  formulation }
\label{DLI_BSplines_MathForm}
In Paper I we gave a detailed description of the mathematical formulation of our algorithm for the SSJE. For completeness we repeat some very basic mathematical formulas that are related to our needs for the present analysis. The interested reader should consult Paper I and references therein for a complete understanding of our method. 
\subsection{B-spline Functions}
\label{DLI_paperI_BSpline_functions}
A B-spline function of order $k$ is the linear combination of some constant coefficients $a_i$ with the B-spline basis functions $B_{i,k}(x)$:   
\begin{equation}
f(x) = \sum_{i=1}^{n} a_i B_{i,k}(x)
\end{equation}  
The $B_{i,k}(x)$ basis functions are known polynomial pieces of degree $k-1$, joined together in a special way as to ensure certain differentiability and smoothness criteria. These B-spline basis functions $B_{i,k}(x)$ are defined over a non-decreasing interval $\xi_0 \leq \xi_1 \leq \cdots \leq \xi_m$ that we call the knot sequence. Each of the $\xi_i$ is called a knot. The distribution of knots $\xi_i$ and the constant coefficients $a_i$ regulate the geometric shape of $f(x)$. For a detailed description of B-spline functions the reader should consult Paper I and references therein. 

\subsection{The Spherically Symmetric Jeans  Equation} \label{DLI_BSplines_Jeans_section}
In Paper I we demonstrated that if we represent the radial second order velocity moment, $\sigma_{rr}^2$,  in a B-spline basis defined over $[0,r_t]$, i.e.: 
\begin{equation}\label{DLI_BSplines_BasisExpansion}
\sigma_{rr}^2(r) = \sum_{i=1}^{N_{\text{coeffs}}} a_i B_{i,k}(r)
\end{equation}
then the line-of-sight velocity dispersion can be written in the form: 
\begin{equation} \label{DLI_BSplines_fundamental}
\sigma_{los}^2(R)  = \sum_i a_i I_i(R) + C(R)
\end{equation}  
where we defined: 
\begin{align} \label{DLI_BSplines_I}
I_{i} (R) &\equiv \frac{1}{\Sigma(R)}
\int_{R}^{r_t}\frac{\left(2 r \rho + \rho^{(1) }R^2\right)  B_{i,k}(r) 
+ \rho  R^2 B_{i,k}^{(1)}(r)  }{\sqrt{r^2-R^2}}\\
\label{DLI_BSplines_C}
C(R) &\equiv\frac{1}{\Sigma(R)}\int_{R}^{r_t} \frac{\rho(r) R^2}{\sqrt{r^2-R^2}} \frac{d \Phi(r)}{dr}
\end{align}
where $r_t$ is the tidal radius of the system.  
Quantities
$\rho^{(1)}(r) \equiv d \rho(r)/dr$ and $B^{(1)}_{i,k}(r) \equiv d B_{i,k}(r) / dr$ represent the first derivative of the mass density and the B-spline basis function.  
Coefficients $a_i$ define the shape of $\sigma_{rr}^2$ and  affect the geometric shape of $\sigma_{los}^2$.

For a system that consists of both stellar and dark matter populations, the above formalism is generalized with the assumption that the mass density $\rho(r)$ is now the tracer stellar density, i.e.  $\rho(r) \to \rho_{\star}(r)$ and   
$\Sigma(R) \to \Sigma_{\star}(R)$. 
The interaction of the stellar with the dark matter component is performed through the total potential 
\begin{equation}
\frac{d\Phi_{\text{tot}}(r) }{dr} = 
\frac{G M_{\text{tot}}}{r^2} = \frac{G (M_{\star} + M_{\bullet})}{r^2}
\end{equation}
where symbol $\star$ corresponds to stellar population, and $\bullet$ to dark matter. 

\subsection{Dynamical Models}\label{DLI_BSplines_DynModel}

In the following sections we will reconstruct, from synthetic data, the kinematic profile. i.e. $\sigma_{rr}^2$ and $\sigma_{los}^2$ (once $\sigma_{rr}^2$ is known, $\sigma_{tt}^2$ can be found from the SSJE), of a stellar system in equilibrium. We will assume that the stellar mass content of this system is described from a  \cite{1966AJ.....71...64K} mass density $\rho(r)$. This is the only stellar density  we are going to consider. For cases where we will consider also a cold dark matter component, we will use a Navarro, Frenk and White  profile   \cite[][hereafter NFW]{1996ApJ...462..563N}. 
The quantity $\sigma_{rr}^2$ is going to be reconstructed from a B-spline function approximation, as discussed in previous sections; in Paper I we gave a detailed description of the necessary functions that we use for our modelling. In short, a King mass model is defined through the parameters $(w_0, \rho_0, r_c)$, where $w_0$ is the value of the transformed potential 
$w(r) = -\Psi(r)/\sigma^2$ at the center $(r=0)$, $\rho_0$ is the stellar core density of the cluster, and $r_c$ the King core radius; the interested reader should consult Paper I and references therein for more details. 

The NFW profile reference is defined through the mass density: 
\[
\rho_{\bullet}(r)= \frac{r_s^3 \rho_{0\bullet}}{r (r_s^2+r^2)^2}. 
\]
where $\rho_{0\bullet}$ is a characteristic  dark matter (hereafter DM)  density  and $r_s$ a characteristic length. 

\section{Statistical Analysis} \label{DLI_BSplines_StatAnal}
In this section we will be using standard  frequentist and Bayesian approaches to model fitting. We will interchange the use of these, according to what best suits our needs each time. Eventually our model fits are going to be performed in a fully Bayesian context. The reader is directed to standard texts such as \cite{hastie01statisticallearning, sivia2006data} and \cite{2010blda.book.....G}  for further details. 

\subsection{Likelihood function}

In this section we augment the formalism we developed in Paper I in order to account also for a smoothing penalty. The full data set of brightness, $D_B$, and kinematics, $D_K$    is $D=\{D_B, D_K \}$. 
The full posterior probability of our complete data set, including the penalty, is:
\begin{equation}\label{DLI_BSplines_posterior}
P(\theta | D)\propto P(\theta) \mathcal{L}(D|\theta) p(W|\lambda) p(\lambda)
\end{equation}
where $\theta$ represents the vector of parameters needed to fully 
describe a given assumed physical model; the terms 
 $P(\theta)$ and  
$\mathcal{L}(D|\theta)=\mathcal{L}(D_B|\theta) \mathcal{L}(D_K|\theta)$
have the same mathematical form as in Paper I. 
The term $p(W|\lambda)$ represents a smoothness\footnote{See following section for definition of smoothness.} penalty that we must apply to the shape of the $\sigma_{los}^2$ function such as to avoid variational behaviour. This is encoded in the $W$ quantity which we shall define later.   
The $p(W|\lambda)$ penalty is governed by the set of values of the $\lambda$ smoothing parameter. $p(\lambda)$ represents the distribution from which our smoothing parameter $\lambda$ draws values and that we are going to determine later. 
We use the same Markov Chain Monte Carlo (MCMC) algorithm as in Paper I.

\subsection{Bayesian Model Selection}
In Paper I we described the necessary mathematical framework  for Bayesian model inference.  Model comparison is performed through the evaluation of Bayesian evidence $Z$. 
For this we use MultiNest \citep{2008MNRAS.384..449F,2009MNRAS.398.1601F}. In cases where the model complexity becomes too great, we are going to use the Bayesian Information Criterion (BIC) for the choice of optimum model. BIC is defined as: 
\begin{equation}
\text{BIC} = - 2 \ln (\mathcal{L}(D|\hat{\theta})) + d \log(N) 
\end{equation}
where $d = \dim (\theta)$ is the total number of free parameters, $N$ is the total number of data points, and $\mathcal{L}(D|\hat{\theta})$ the maximum likelihood value as results from an MCMC run.  $\hat{\theta}$ are the values of parameters, $\theta$,  that maximize the likelihood function, $\mathcal{L}(D|\hat{\theta})$.
We can legitimately use BIC, since the posterior distribution belongs to the exponential family. When we perform model comparison using BIC, the most probable model is the one with smallest BIC value.

\subsection{Smoothing Penalty Distributions}

In this section we are going to describe our choice of probability distributions $p(W|\lambda)$ and $p(\lambda)$. 
By the term ``smoothing'' we mean how ``stiff'' or ``non-flexible'' our resulting B-spline function must be. The probability distributions $p(W|\lambda)$ and $p(\lambda)$ govern this property of the resulting $\sigma_{los}^2$ function as well as of the $\psi(r)=\sigma_{rr}^2(r)$ B-spline representation. 
We need smoothing penalty for cases where we have large errors and the B-spline function tends to follow the data in an unphysical way. We also need this penalty  for cases of incomplete data: they do not span all distance $[0,r_t]$).

For the penalty function $W$ we choose the following scheme: 
\begin{equation}
W = \int_0^{r_t} \left[ q \left(\frac{d^2 \sigma_{los}^2}{dR^2}\right)^2 + (1-q) \left(\frac{d \sigma_{los}^2}{dR} \right)^2
\right] dR.
\end{equation}
We penalise both the first (slope) and second (curvature) derivatives of the line-of-sight velocity dispersion. We choose to penalise   $\sigma_{los}^2(R)$ and not $\psi=\sigma_{rr}^2(r)$ since the former is compared directly to the data set.  
The parameter $q$ lies in the range $q \in [0,1]$  and describes at what percentage each of the derivatives of $\sigma_{los}^2$ participates in the penalty. Our choice, after a lot of trial and error, will be fixed\footnote{We finalised this value on experiments with synthetic data and known models, by calculating their Bayesian evidence.} at $q=0.25$. That is, we assign the majority of the penalty to the second derivative, thus penalising mainly the curvature of the curve, and not the slope.

In a full Bayesian context we have: 
\begin{equation}\label{DLI_BSplines_penalty_pdf}
p(W|\lambda) = \lambda e^{-\lambda W}
\end{equation}  
which is normalised  in the  range of all positive values $W \in [0,\infty)$:
\begin{equation}
\int_0^{\infty} \lambda e^{-\lambda W} dW = 1.
\end{equation} 
$p(\lambda)$ is going to be another  ``prior'' distribution of values over $\lambda$. In statistician's terminology, this is called a hyperprior. 
We will assume that $p(\lambda)$  has the form of the inverse Gamma distribution\footnote{The motivation of our choice will be described in the next section.}, i.e.:
\begin{equation}\label{DLI_BSplines_IGprior}
p(\lambda) \equiv p(\lambda|\alpha,\beta) = 
\frac{\beta^{\alpha}}{\Gamma(\alpha)}
\lambda^{-\alpha-1} e^{-\frac{\beta}{\lambda}}
\end{equation}  
Based on the above definitions we see that a large value of $\lambda$ tends to make $\sigma_{los}^2$ a straight line. On the contrary a very small value allows the model to overfit.

We have to note that instead of the combination $\lambda$ and fixed $q$ we could use two different free parameters, $\lambda_1$ and $\lambda_2$, each penalising a different derivative of $\sigma_{los}^2$. However this would increase the complexity of our model and our calculations, with a small extra gain. For a full discussion on penalties on smoothing functions the interested reader should consult \cite{hastie01statisticallearning}. Again we note that in the smoothing splines fitting, from a statistician's perspective, all penalties are applied to the smoothing spline. Here we follow a significantly different approach, penalising a function which is the end product of some very complex calculations. That is, we apply the smoothing penalty to the function that is directly fitted to observables and not to the smoothing B-spline function.

\subsection{Training the model for Optimum Smoothing}
\label{DLI_BSplines_training_optimum_smoothing}
In this section we are going to describe how we estimate the parameters $\alpha$ and 
$\beta$ of the $p(\lambda|\alpha,\beta)$ distribution for optimum smoothing for cases where we consider penalty in our models. Once we have  $p(\lambda|\alpha,\beta)$ defined, we can use $\lambda$ as a free parameter in the MCMC walk. Our reasoning is that  different models must behave in a similar way in terms of their ``stiffness'' behaviour (smoothing). That is, we expect, in terms of smoothness only, that a real physical system will approximately have the same behaviour as an ideal theoretical model with a known kinematic distribution profile. Then we are going to use theoretical known models, in order to have a measure of smoothness that is needed when we fit our model to  real data.

\label{DLI_BSplines_MachineLearning}
\begin{figure}
\centering
%height=0.3\textwidth, width=0.5\textwidth
%\showthe\columnwidth 
\includegraphics[width=\columnwidth]{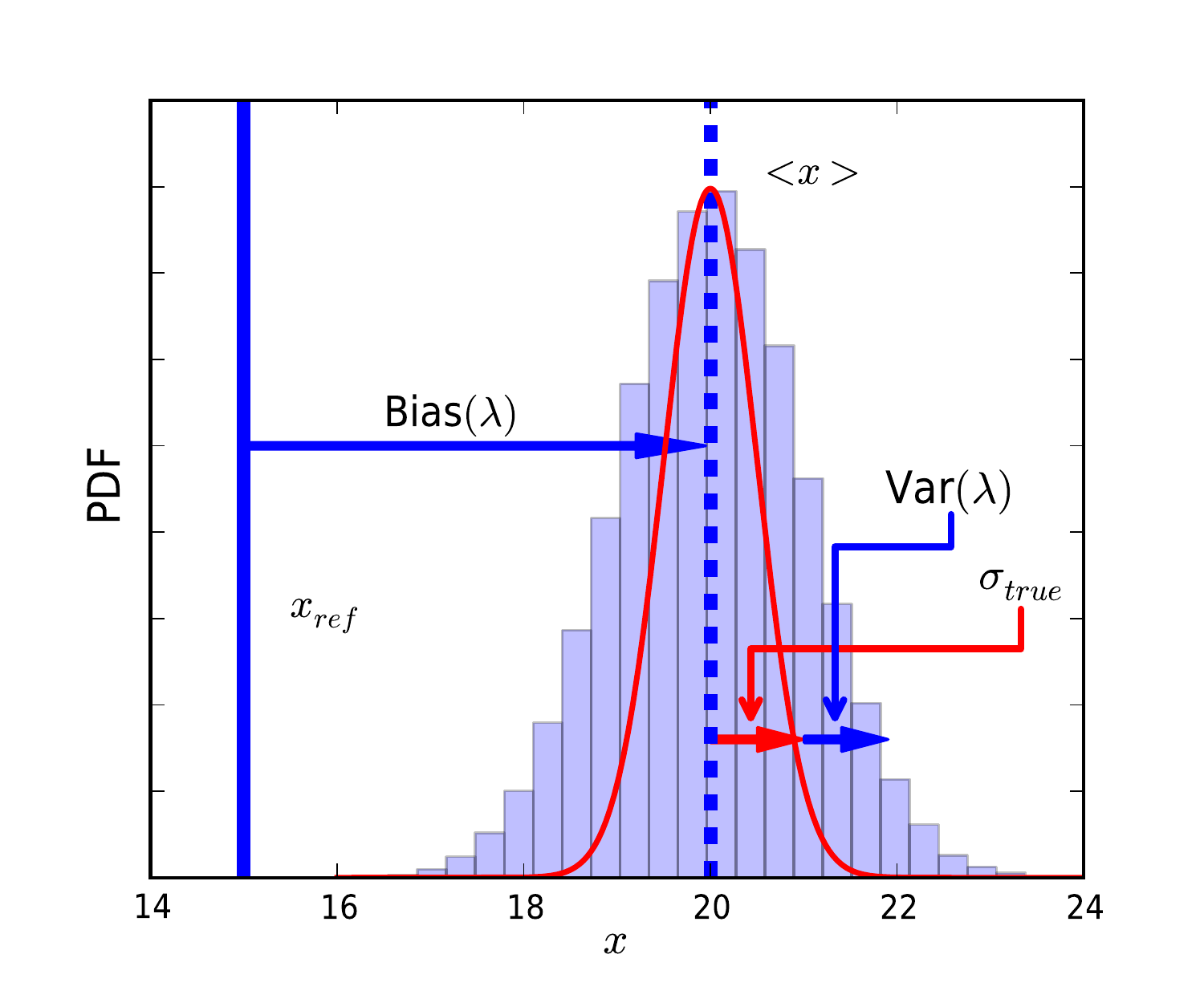}
 \caption{ The histogram represents the distribution of values from an MCMC walk that have a biased mean estimate, and a biased variance. 
   The mean value $\langle x \rangle \sim 20$ differs from the correct 
   $x_{\text{ref}} \sim 15$ by the amount: $\bias  =  \langle x \rangle - x_{\text{ref}} $. The variance 
   $\sigma^2$ of the histogram is increased by an amount $\Var(\lambda)$ according to  $\sigma^2=\sigma_{\text{true}}^2+\Var(\lambda)$. }
\label{DLI_BSplines_bias_hist}
\end{figure}

We must be careful in the choice of the parameters $\alpha, \beta$ of $p(\lambda|\alpha, \beta)$. A ``stiff'' curve may result in underfitting, while 
alternatively a very ``soft'' curve can result in overfitting. 
This can also be affected by the number of coefficients $a_i$. 
The fewer their number, the less flexible the resulting B-spline function. 
We must draw attention to the fact that there does not exist a single range of values for $\lambda$ for all data sets. This range depends  on the number of available data, the choices we make for the knot distribution (uniform, exponential etc) and the number of coefficients. 
Especially a large number of data points, has a severe impact on the resulting value of $\mathcal{L}(D|\theta)$ since the  likelihood product, $\mathcal{L}(D_B|\theta) \mathcal{L}(D_K|\theta)$ (see Paper I), contains more terms, each smaller than unity. 
Then the quantities $p(W|\lambda)$ and $p(\lambda|\alpha,\beta)$ must take a different range of values in order not to have a larger (or smaller) effect on the resulting likelihood value. This is regulated by parameters $(\alpha,\beta)$ of $p(\lambda|\alpha,\beta)$. 
 For our method to be reliable in the case where we use a penalty, we need to know the number of available data and keep fixed in our fittings the choice of knot distribution\footnote{We emphasise that it is the choice of distribution (e.g. uniform, Gaussian etc) that must be fixed, not the knot sequence.}, B-spline order $k$ and number of coefficients. In practice, since the value of $\lambda$ is eventually determined through the MCMC scheme, we can have a small deviation from the above mentioned parameters. This should be avoided for applications to real data sets. 
For example calculate $\alpha, \beta$ for a given number of coefficients and then apply this range to a B-spline function with a   few more coefficients or a few more data points. We will demonstrate with an example, how altering the value of data points and keeping fixed the range of $\lambda$ can result in a small bias in the estimates.

Consider the case where we create a small number of  synthetic data from a model that depends on  a single variable, say $x$. These data are created from a reference value, say $x_{ref}$, in which we add a random Gaussian noise $\mathcal{N}(0,\sigma_{\text{true}})$ of given variance $\sigma_{\text{true}}^2$. We wish to recover this $x_{ref}$ value, within some statistical uncertainty,  through a Markov Chain Monte Carlo scheme. That is, we wish to estimate a marginalised distribution of $x$ values.

Assume that we define a likelihood that depends on $x$ and also on some parameter $\lambda$ that is not estimated from the MCMC. The value of this parameter affects the estimation of $\langle x \rangle$ by inserting some unknown bias in the true mean 
$\langle x \rangle_{\text{true}} = x_{ref}$ and by increasing the total variance  of
the true $x$ values $\sigma^2_{\text{true}}$ by some unknown amount $\Var$. 
Then this bias and variance will be in general a function of $\lambda$, i.e. $\bias \equiv \bias(\lambda)$ and $\Var\equiv \Var(\lambda)$.  The relation between the true reference values of mean and  variance and the estimated  ones is: 
\begin{align}\label{DLI_BSplines_bias}
\langle x \rangle  &= x_{ref} +  \bias(\lambda)\\
\sigma^2 & = \sigma^2_{\text{true}} + \Var(\lambda)
\end{align}
We can see these relations schematically in Fig. \ref{DLI_BSplines_bias_hist}. The histogram represents the estimates of some MCMC walk based on synthetic data points $d_i$. These were created from the true value $x_{ref}$ by adding  Gaussian random noise. Overplotted is a Gaussian (red in color version) from which we generated a synthetic data sample of values $x_i$, centred on the (biased) mean value  $\langle x \rangle$.  We show how  the mean $\langle x \rangle$   deviates from  the true $x_{ref}$ due to $\bias(\lambda)$.  We also plot the increase in variance $\sigma^2$ from the true $\sigma^2_{\text{true}}$ due to $\Var(\lambda)$.
\begin{figure*}
\centering
%height=0.3\textwidth, width=0.5\textwidth
%\showthe\columnwidth 
\includegraphics[width=\textwidth]{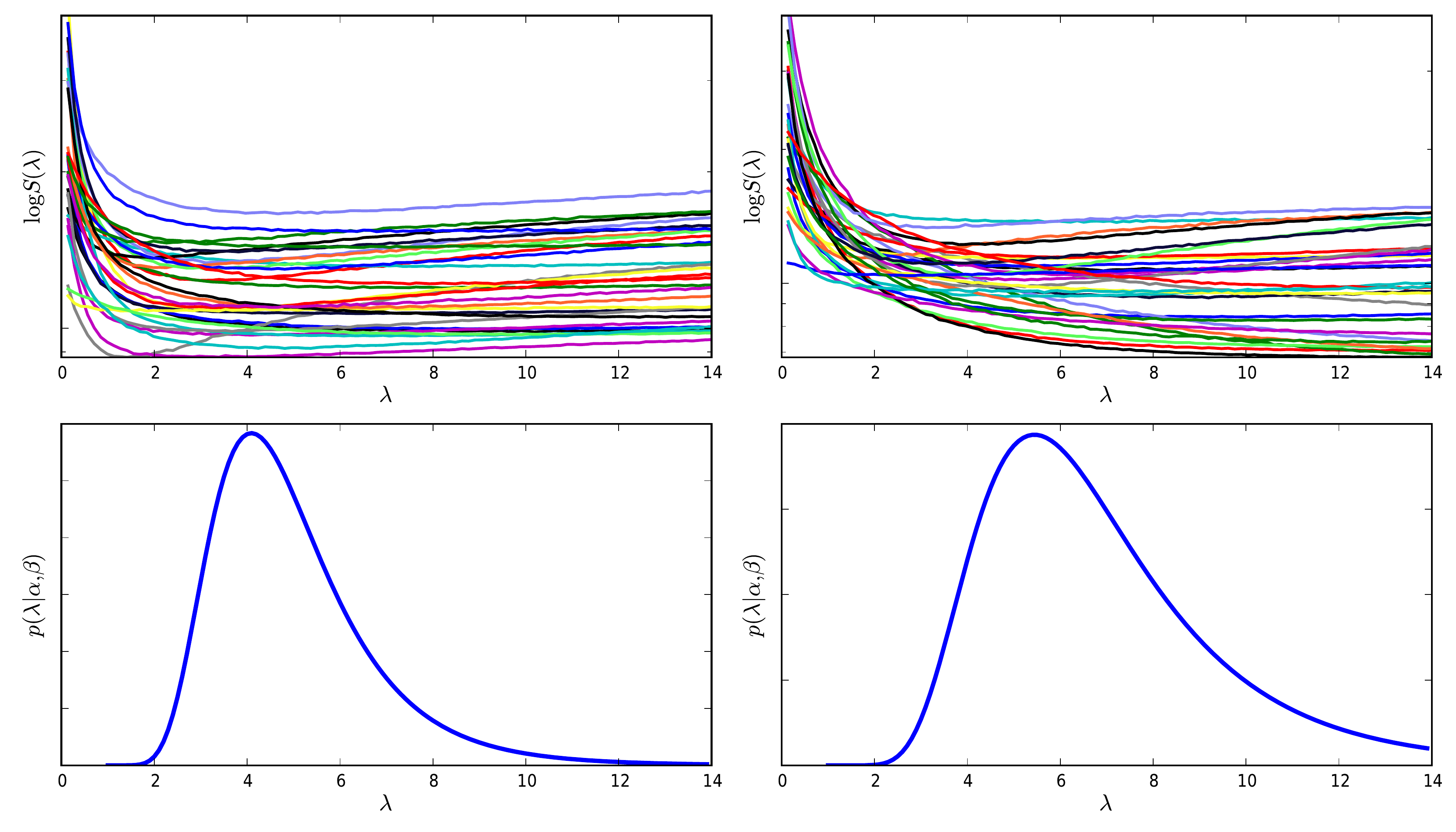}
 \caption{ We plot the value of the logarithm of the sum $S(\lambda)$ for 30 randomly created independent data samples of synthetic brightness $J$ and line-of-sight velocity dispersion $\sigma_{los}^2$. Top left panel: $\log S(\lambda)$ from random samples  created from a King mass model with Osipkov-Merritt anisotropy profile. Bottom left panel: Inverse Gamma distribution, for the optimum choice of $(\alpha, \beta)=(10,45)$ parameters, as estimated from MCMC scheme. Top right panel: $S(\lambda)$ from random samples  created from an Isotropic King model. 
 Bottom right panel: Inverse Gamma distribution, for the optimum choice of 
$(\alpha, \beta)=(8,49)$  parameters, as estimated from MCMC scheme.  
}
\label{DLI_BSplines_Shat}
\end{figure*}

We wish to find which is the optimum $\lambda$ value that gives the optimum 
trade-off between bias and variance in our estimates. We define as measure of this trade-off  the average deviation of each of the MCMC walk points $x_i$ from the reference value $x_{ref}$: 
\begin{equation}\label{DLI_BSplines_hatS}
S(\lambda) = \frac{1}{N_{\text{MCMC}}} \sum_i^{N_{\text{MCMC}}} (x_i-x_{ref})^2
\end{equation}
Substituting Equation \ref{DLI_BSplines_bias} in \ref{DLI_BSplines_hatS} yields:
\begin{multline*}
S(\lambda) = \frac{1}{N_{\text{MCMC}}} \sum_i^{N_{\text{MCMC}}}  
\biggl[ 
(x_i - \langle x \rangle)^2   \\
- 2 (x_i - \langle x \rangle) \bias(\lambda) + \bias^2(\lambda)
\biggr],
\end{multline*}
that is
\begin{equation} \label{DLI_BSplines_hatS_reduced}
S(\lambda)=  \sigma^2(\lambda) + \bias^2(\lambda),
\end{equation}
where $\bias^2(\lambda) = \left[\bias(\lambda)\right]^2$, and we used: 
\[
 \frac{1}{N_{\text{MCMC}}} \sum_i^{N_{\text{MCMC}}}  (x_i - \langle x \rangle) \bias(\lambda)=0
\]
In Equation \ref{DLI_BSplines_hatS_reduced}, $\sigma^2(\lambda)$  can be further decomposed into the true variance $\sigma_{\text{true}}^2$ of the target distribution around its mean, that is inherent to  the process 
by which we created the random sample, and a variance term $\Var(\lambda)$ that depends on the smoothing parameter\footnote{In general we expect that a ``stiff'' curve will have smaller variation in its estimates when fitting to a set of data. Quantity $\Var(\lambda)$ expresses exactly this fact. } $\lambda$: 
\begin{equation} \label{DLI_BSplines_fittness_measure}
S(\lambda)=  \sigma_{\text{true}}^2 + \Var(\lambda) + \bias^2(\lambda),
\end{equation}
That is, the deviation of the values $x_i$ of the MCMC chains from the reference value,  is some irreducible variance $\sigma_{\text{true}}^2$ that is inherent to the random  process in which the data were created, and a part that results from the additional bias $\bias(\lambda)$ and variance $\Var(\lambda)$ inserted by a wrong assumption on the value $\lambda$. 
 There is an interplay between $\Var(\lambda)$ and $\bias(\lambda)$. As we  increase $\lambda$ parameter from zero, we expect the term $\Var(\lambda)+ \bias^2(\lambda)$ to be reduced. However, beyond some value, 
while $\Var(\lambda)$ will continue decreasing, $\bias(\lambda)$ will start rising and result in an overall increase\footnote{  \citep[For a full discussion on the Bias - Variance trade off see][]{hastie01statisticallearning}.} in $S(\lambda)$. Therefore, the optimum value of $\lambda$ parameter is going to be the one that minimises $S(\lambda)$
 in the estimates from the MCMC.

We are going to use  this ``variance'' $S(\lambda)$ as a measure of optimum smoothing for the case of a B-spline representation of $\sigma_{rr}^2$. For this case, we define: 
\begin{equation}\label{DLI_BSplines_S}
S(\lambda) =  \frac{1}{N_{\text{MCMC}}}  \sum_j^{N_{\text{MCMC}}} 
\left[\frac{1}{n} \sum_{i}^{n}(a_i-a_i^{ref})^2 \right]_j.
\end{equation}
The quantity inside the brackets corresponds to the average  deviation $(a_i-a_i^{ref})^2$ of all the coefficients $a_i$ from their corresponding reference value, 
for each proposed value from the MCMC walk.  
We estimate each $a_i^{ref}$ reference value that corresponds to the reference radial velocity dispersion $\sigma_{rr}^2(r)|_{ref}$ of a known analytic model, through: 
\begin{equation} \label{DLI_BSplines_sigma_ref_eq}
\sigma_{rr}^2(r)\bigg|_{ref} = \sum_i a_i^{ref} B_{i,k}(r) 
\end{equation}
by convolving Equation \ref{DLI_BSplines_sigma_ref_eq} with $B_{j,k}(r)$, i.e.: 
\begin{equation}  \label{DLI_BSplines_exactSolution1}
\int_0^{r_t}\sigma_{rr}^2(r)\bigg|_{ref}  B_{j,k}(r) dr= \sum_i a^{ref}_i \int_0^{r_t} B_{i,k} (r) 
B_{j,k} (r) dr 
 \end{equation}
 where $j=0,\ldots,n$. 
 This results in a system of $n$ equations for the $n$ unknowns $a_i^{ref}$ which can be solved with standard linear algebra methods.

\subsubsection{Determination of $\alpha, \beta$ parameters of the $p(\lambda|\alpha,\beta)$ distribution.}
\label{DLI_BSplines_ML_detail}

For model fits where we will use a penalty function, we choose to use seven\footnote{We remind  the reader that the last index $a_n=0$ since  at the tidal radius of the system $\psi(r_t)=\sigma_{rr}^2(r_t)=0=a_n$. This follows from the fact that the B-spline represented function passes through the last point, see section $2.2$. So the last coefficient $a_8$ is going to be fixed.} 
 $a_i$ unknown coefficients, i.e. $n=7+1$ and a uniform distribution of knot values. The order of the B-spline representation is set to $k=5$. 

The first thing we need is to  determine a range of values $\lambda$ that give optimum penalty. That is, create a set of $\lambda^j_{\text{est}}$ estimated $\lambda$ values that minimise the ``measure of fitness'' $S(\lambda)$.
 Then, using these observed values as ``data'' we need to estimate the values of $\alpha, \beta$ that go into the hyperprior $p(\lambda|\alpha, \beta)$ of the full posterior distribution (Equation \ref{DLI_BSplines_posterior}). 
 
 For the determination of a range of values based on the measure of optimum smoothing (Equation \ref{DLI_BSplines_S}) we use as likelihood $\tilde{\mathcal{L}}$ $(\neq \mathcal{L})$ the functional form: 
\begin{align}
\tilde{\mathcal{L}}(\theta | D) &\propto  \mathcal{L}(D|\theta) p(W|\lambda) \\
\tilde{\mathcal{L}}(\theta | D) &\propto  \mathcal{L}_B \mathcal{L}_K
\lambda e^{-\lambda W}
\end{align}
i.e. we removed the unknown hyperprior distribution $p(\lambda|\mu_{\lambda},\sigma_{\lambda})$ and the prior probability $p(\theta)$. The former is not needed, since we are going to use a fixed $\lambda$ for each MCMC evaluation, while the latter does not affect our calculations (it is constant and is simplified out in the MCMC process).

%The algorithm we  use can be seen in Algorithm \ref{DLI_BSplines_ML_alg}.

The steps of our algorithm in detail are:
\begin{enumerate}
\item Create a large number of  training data sets, $N_{\text{Sample}}$ from a known theoretical model. .
\item For each of these data sets: 
\begin{enumerate}
\item For the range of $\lambda$ values $[0,\lambda_{max}]$, keeping the mass model fixed, estimate the 
marginalized distributions of the $a_i$ coefficients. 
\item Find the approximate value of $\lambda$  that minimizes  the ``measure of fitness'' 
$S(\lambda)$ of the produced MCMC chains from the ideal model and store it as $\lambda^j_{\text{est}}$.  
\end{enumerate}
\end{enumerate}

%%%%%%%%%%%%%%%%%%%%%%%%%%%%%%%%%%%%
\begin{comment}
\IncMargin{1em}
\begin{algorithm} 
\SetKwData{Run}{run}
\SetKwFunction{Create}{Create}
\SetKwFunction{Evaluate}{Evaluate}
\SetKwData{Increase}{Increase}
\SetKwData{Up}{up}
\SetKwFunction{Union}{Union}
\SetKwFunction{Find}{Find}
\SetKwInOut{Input}{input}\SetKwInOut{Output}{output}
\DontPrintSemicolon
%\SetAlgoLined
%\Input{A bitmap $Im$ of size $w\times l$}
%\Output{A partition of the bitmap}
\BlankLine
\For{$j \leftarrow  1$ \KwTo $N_{\text{Sample}}$}{
\BlankLine
\Create Random Data set\;
\BlankLine
\For{$\lambda \leftarrow 0$  \KwTo $\lambda_{max}$}{
\BlankLine
\Run MCMC\;
\Evaluate $S(\lambda)$\;
\Increase $\lambda$ by a small step $\delta \lambda$\;
}
\Find $\hat{\lambda}$ that minimizes $S(\lambda)$ and store as  $\lambda^j_{\text{est}}$
}
\caption{Creation of $\lambda_{\text{est}}$ sample\label{DLI_BSplines_ML_alg}}
\end{algorithm}
 \end{comment}
%%%%%%%%%%%%%%%%%%%%%%%%%%%%%%%%%%%% 
 
For the determination of $\alpha, \beta$ values, we assume that our estimate  $\lambda_{\text{est}}$, from each random sample, has a Gaussian random  error deviation from the true value $\lambda_{\text{true}}$. Then we convolve this Gaussian error with  the hyperprior $p(\lambda_{\text{true}}|\alpha,\beta)$, and integrate over all 
possible values of $\lambda_{\text{true}}$. The resulting function is going to be the distribution function  that $\lambda_{\text{est}}$ values satisfy. We are going to use this distribution function in an MCMC scheme for the definition of the likelihood that eventually will give us the estimates of $(\alpha, \beta)$. 
From each random $j=1,\ldots,N_{\text{Sample}}$ sample of  line-of-sight values $\sigma_{los}^2$ we have a $\lambda^j_{\text{est}}$ value that minimises the $S(\lambda)$ function. That is, we have a  data set $D_{\lambda}=\{\lambda^j_{\text{est}} \}$.  We use this set as ``observable'' in a MCMC algorithm, in order to estimate  parameters $\alpha, \beta$ of the hyperprior $p(\lambda|\alpha,\beta)$. The likelihood we use in this MCMC scheme is the distribution function of $\lambda_{\text{est}}$ values: 
\begin{equation} \notag
\mathcal{L}(D_{\lambda}|\alpha, \beta,\sigma_{\text{est}}) =
\prod_{j=1}^{N_{Sample}}\mathcal{L}(\lambda^j_{\text{est}}|\alpha, \beta,\sigma_{\text{est}})
\end{equation}
where
\begin{multline} \label{DLI_BSplines_IG_lkhood}
\mathcal{L}(\lambda^j_{\text{est}}|\alpha, \beta,\sigma_{\text{est}})=\\ \int_{\lambda^j_{\text{true}}=0}^{\infty} p(
\lambda_{\text{true}}| \alpha, \beta) 
\frac{\exp \left( 
-\frac{(\lambda^j_{\text{est}}-\lambda_{\text{true}})^2}{2 \sigma^2_{\text{est}}}
\right)
}{ \sqrt{2 \pi } \sigma_{\text{est}}}
d \lambda_{\text{true}}
\end{multline}
The output of this MCMC is marginalised distributions of $\sigma_{\text{est}}, \alpha$ and $\beta$. We use the mode values of the $(\alpha, \beta)$  distributions for the hyperprior $p(\lambda|\alpha,\beta)$.

In Fig. \ref{DLI_BSplines_Shat} we plot the optimum smoothing measure $S(\lambda)$ for 
30 
randomly created independent samples.
 These random samples were created from an Isotropic King profile (top right panel) and a King mass model with  Osipkov-Merritt $\beta$ anisotropy profile (top left panel). The value of $\lambda$ that approximately minimises $S(\lambda)$, is somewhere in the interval $[0,14]$. Observe that the lower boundary is steeper than the upper boundary.  Based on this observation, and the fact that the convolution of the Inverse Gamma distribution with a Gaussian is finite, we choose to draw $\lambda$ values from the Inverse Gamma distribution. 
 
 Recall Equation \ref{DLI_BSplines_IGprior} for the assumption on the hyperprior from which the $\lambda$ distribution satisfies.  In bottom panels, we plot the hyperprior $p(\lambda|\alpha,\beta)$ for the optimum values  of $\alpha, \beta$.

\section{Examples} 
\label{DLI_BSplines_Examples}

In this section we are going to reconstruct models of stellar clusters from synthetic data for a variety of anisotropy $\beta(r)$ functions. For the notation of the total number $n$ of unknown coefficients $a_i$, as mentioned earlier,  we use the following scheme: Based on the restriction that all the quantities that describe the cluster must be zero at the tidal radius, the final coefficient will be $a_{n}=0$. This extra coefficient does not go into the likelihood analysis, hence we break the total number of coefficients to the sum of unknowns plus one which represents this last coefficient. We use this notation in captions of Figures and in Tables where we list Bayesian evidence or BIC values.

 Our motivation is as follows: 
in general the King brightness profile gives a very good fit to observational data. 
Thus, it can be a  good approximation to the mass model. However the kinematic profile is not always adequate and can affect the mass measurements.
 So  we allow information from the line-of-sight velocity dispersion data to accurately reconstruct the shape of the B-spline representation of $\sigma_{rr}^2$ and consequently $\sigma_{los}^2$. Eventually
the kinematic profile and  the mass  content of the cluster is fully reconstructed independently of the anisotropy of the system. This independence results from the flexibility of the smoothing spline representation of    $\sigma_{rr}^2$. 
We will see that we recover accurately in all cases the correct reference parameters of the mass model and the mass-to-light ratio and to a good accuracy the second moments of the radial and line-of-sight velocities. 
We find no degeneracy in our results, i.e. a unique kinematic profile is constructed for each example. Finally, the quality of our fits depends on the available data set.

\begin{figure}
\centering
%height=0.3\textwidth, width=0.5\textwidth
%\showthe\columnwidth 
\includegraphics[width=\columnwidth]{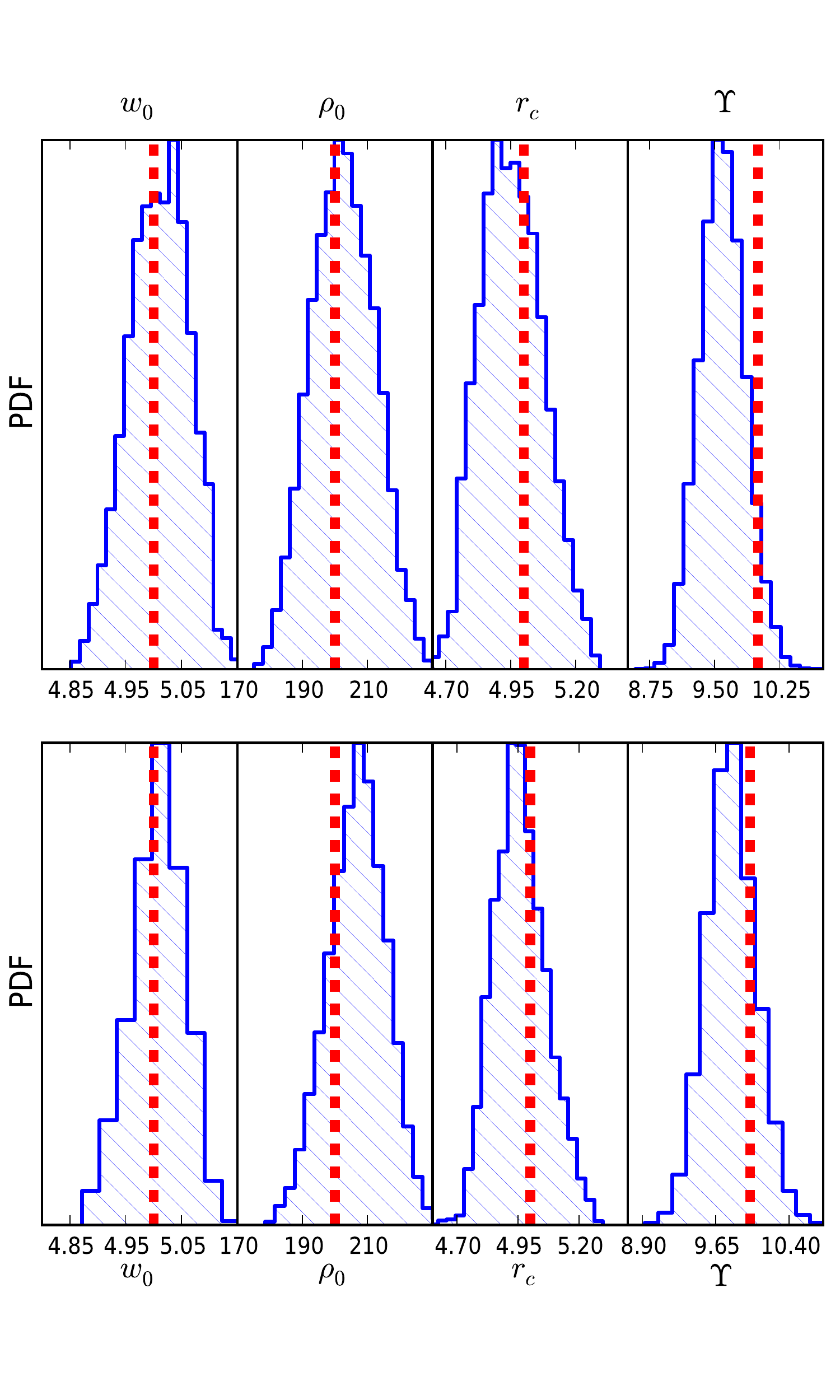}
 \caption{Histograms for the system with Osipkov-Merritt anisotropy $\beta_1 = Q \frac{r^2}{rc^2+r^2}$, with $Q=0.25$ and $r_c$ the King core radius. The red dashed lines correspond to the reference values $\{w_0^{ref},\rho_0^{ref},r_c^{ref},\Upsilon^{ref}\}=\{5,200,5,10\}$ from which synthetic data were created. Top panel: marginalised distributions of the mass model  parameters $(w_0,\rho_0,r_c,\Upsilon)$ for $n=5+1$ coefficients $a_i$ model, with no penalty. Bottom panel:  
 The same mass model parameters, but now for the case of $n=7+1$ coefficients $a_i$ with  smoothing penalty.  For both cases the reference values of the mass model are well within the boundaries of the estimated values. }
\label{DLI_BSplines_FigJeans_params}
\end{figure}

\begin{figure*}
\centering
%height=0.3\textwidth, width=0.5\textwidth
%\showthe\columnwidth 
\includegraphics[width=\textwidth]{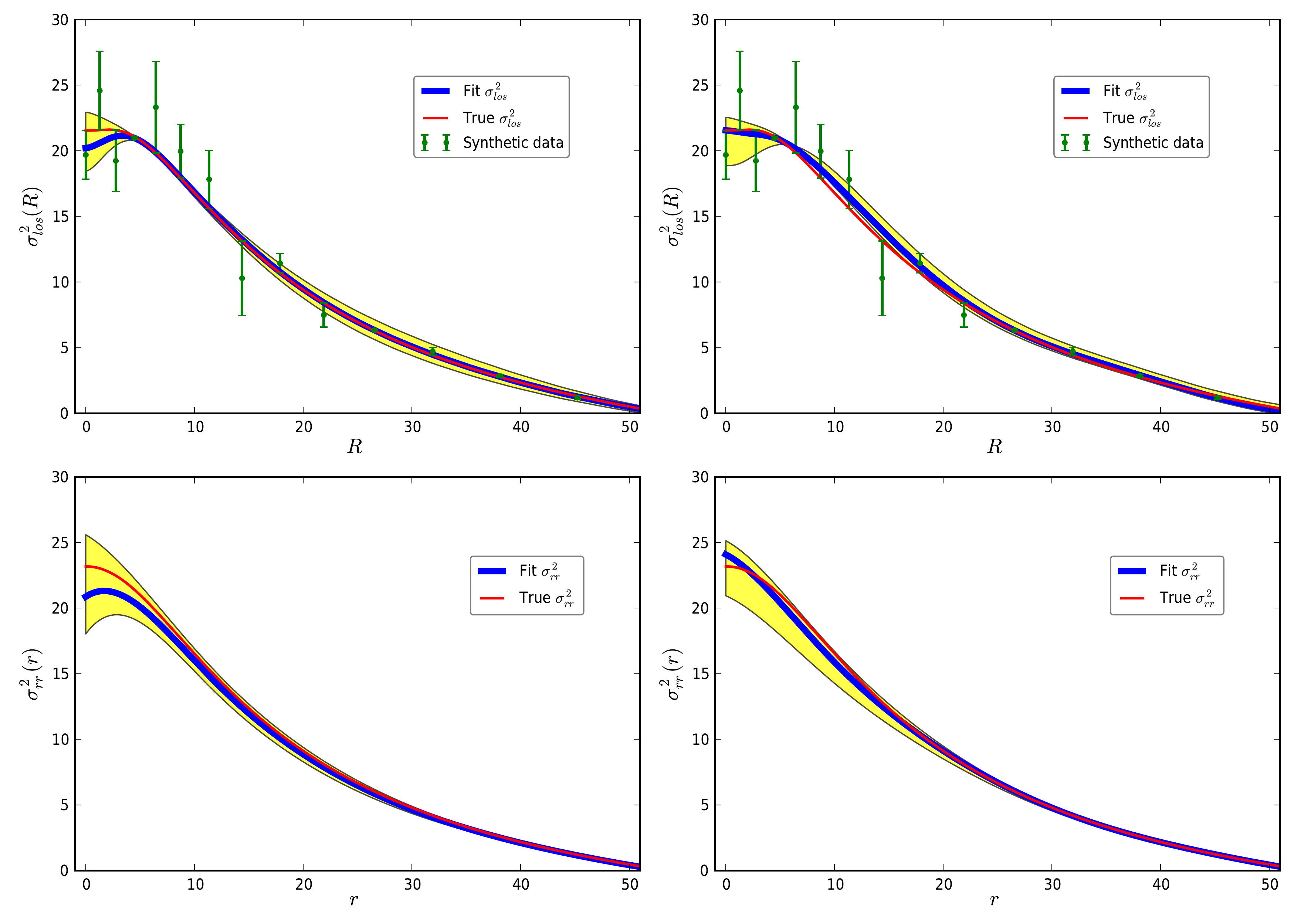}
 \caption{King mass model with Osipkov-Merritt anisotropy. Left panels correspond to $n=5+1$   coefficients $a_i$.    Top left panel:  synthetic data, true and highest likelihood  fit  of $\sigma_{los}^2$.  Bottom left panel: corresponding $\sigma_{rr}^2$ fit and reference values.   
 Yellow shaded regions correspond to the  $1\sigma$ uncertainty interval of the unknown coefficients $a_i$ only, i.e. we do not account for the uncertainty of the mass defining parameters $(w_0,\rho_0,rc)$. 
Righthand panels correspond to the same mass model and anisotropy, however now we consider $n=7+1$ coefficients $a_i$ and the full posterior distribution with penalty $p(W|\lambda)$ and hyperprior $p(\lambda|\alpha,\beta)$ (Equations \ref{DLI_BSplines_penalty_pdf} and \ref{DLI_BSplines_IGprior}). 
 }
\label{DLI_BSplines_FigJeans_OM_fits_data}
\end{figure*}

\begin{table}
\caption{Bayesian evidence for Osipkov-Merritt anisotropy and King mass density.}
\label{DLI_BSplines_Example1OM_14_evidence}
\centering
\begin{tabular}{@{} c|  c | c | c }
 \hline 
Penalty & NDim & Number of coefficients $n$  & $\ln (Z) \pm \delta \ln Z$  
 \\ \hline \hline
No & 9  & $n=5+1$     & $ -51.82 \pm  0.19$  \\[6pt]  \hline 
No & 10  & $n=6+1$     & $-52.86\pm  0.19$  \\[6pt]  \hline 
 Yes & 12 &  $n=7+1$   & $-68.68 \pm  0.20$  \\[6pt] 
\hline
 Yes & 13 &  $n=8+1$   & $ -71.58\pm 0.21 $  \\[6pt] 
\hline
\end{tabular}
\medskip\\
The order of the B-spline representation of $\psi\equiv \sigma_{rr}^2(r)$ was held fixed at $k=5$. 
From left to right: first column describes if we used penalty or not. Second column is the total dimensionality of the model.   Third column is the number of coefficients $a_i$ and the fourth column is the value of Bayesian evidence. The highest value of $\ln Z$ corresponds to the most probable model.  
\end{table}

\subsection{Osipkov-Merritt anisotropy and King mass density}

In this example we consider a King mass model and a kinematic profile constructed from an Osipkov-Merritt anisotropy $\beta(r)$ function of the form: 
\begin{equation}
\beta(r) = Q \frac{r^2}{r_c^2+r^2}.
\end{equation}
For the creation of the synthetic data points we use the same process as in 
Paper I. 
In this example the mass-to-light ratio $\Upsilon$ is a free parameter. The reference values for the creation of synthetic data we use are $\{ w_0^{ref}=5, \; \rho_0^{ref} = 200,\;  r_c^{ref} = 5,\;  \Upsilon^{ref}=10,\;  Q=0.25\}$. 
We use 14 synthetic data points and an increased error of  $20\%$ of the true value.  For the B-spline representation of $\psi\equiv\sigma_{rr}^2(r)$ we use uniform knot distribution in the interval $[0,r_t]$. Observe that this knot distribution is adaptive: for each proposed value of parameters $w_0,\rho_0,r_c$, the solution of Poisson Equation yields a tidal radius $r_t$. For this $r_t$ we construct a uniform knot distribution. This is repeated for each likelihood call in the MCMC.

First we fit our model to the synthetic data with no penalty. In this case we use 
$n=5+1$ coefficients, $a_i$. Then we fit including a  penalty distribution as described in section \ref{DLI_BSplines_MachineLearning} and using $n=7+1$ coefficients, $a_i$. 
The values of  $\alpha,\beta$ in the hyperprior $p(\lambda|\alpha,\beta)$ are the ones that correspond to the estimates from the King Isotropic model (section \ref{DLI_BSplines_ML_detail}). We use a higher number of coefficients $a_i$ in the case where we also consider a penalty in order  to avoid a very ``stiff'' $\sigma_{los}^2$ function and give greater fitting flexibility to our model. The parameters $\alpha,\beta$ of the hyperprior were calculated for $n=7+1$ coefficients $a_i$; we list the values of Bayesian evidence in Table \ref{DLI_BSplines_Example1OM_14_evidence}. The models that are favored from Bayesian evidence are the $n=5+1$ for the case with no penalty, and $n=7+1$ for the case with penalty.

In Fig. \ref{DLI_BSplines_FigJeans_params} we plot the histograms of the mass model parameters for the two fits;  the top panel corresponds to the fit with no penalty, while the bottom panel is the fit with penalty. In both cases the reference values (vertical dashed curves) from which the synthetic data were created are within the uncertainty range of the marginalised distributions; the mass content of the model is fully reconstructed.

In Fig. \ref{DLI_BSplines_FigJeans_OM_fits_data} we plot the fits for the second moments of the line-of-sight and  radial  velocities; while left panels correspond to the fits with no penalty considered. 
In all panels the yellow shaded regions correspond to the  $1\sigma$ uncertainty interval of the unknown coefficients $a_i$ only, i.e. we do not account for the uncertainty of the mass defining parameters $(w_0,\rho_0,rc)$.  
Right panels correspond to fits with penalty;  note that the fit takes into account the brightness synthetic profile as well, although we do not plot it here. 
Top left panel is the line-of-sight velocity dispersion, $\sigma_{los}^2$, and the 14 synthetic data points we employed. We plot the highest likelihood 
profile as well as the true function, demonstrating that the fit is excellent. There is a small deviation close to the origin, since  $\sigma_{los}^2$ tends to follow the first data point. However, the uncertainty in the $a_1$ coefficient includes the true value of $\sigma_{los}^2$. The bottom left panel is the corresponding $\sigma_{rr}^2$ as estimated from the line-of-sight fit, although note that we have no data to compare directly to $\sigma_{rr}^2$. This is estimated from the $\sigma_{los}^2$ corresponding fit. Close to the origin there is a deviation of the true 
$\sigma_{rr}^2$ from the estimated one. This again is owing to the tendency of $\sigma_{los}^2$ to follow the first datum, however, the uncertainty interval of $\sigma_{rr}^2$ encapsulates the true value. What is more important is that the mass content is fully reconstructed and is not affected by this small deviation.   
The top right panel is the line-of-sight $\sigma_{los}^2$ fit for the case where we add a smoothness penalty term to the likelihood. In this case the fit is much better, since the first data point does not significantly 
influence the behaviour of the curve.  In the bottom right panel is the corresponding $\sigma_{rr}^2$ fit, and both fits are excellent and closer to the true value than in the case with no penalty.

\begin{figure*}
\centering
%height=0.3\textwidth, width=0.5\textwidth
%\showthe\columnwidth 
\includegraphics[width=\textwidth]{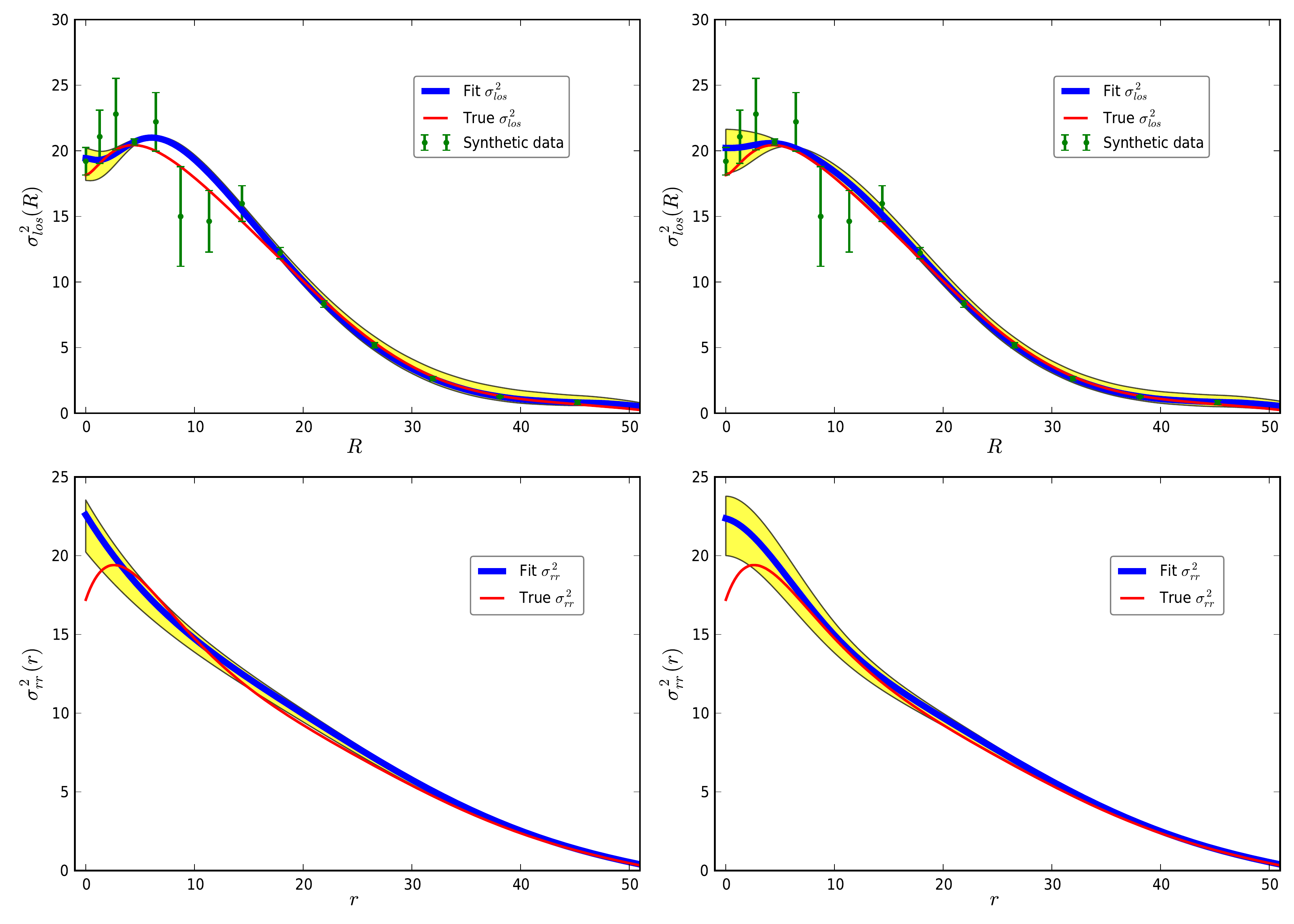}
 \caption{
The fits for the cases of  King mass model with sinusoidal anisotropy parameter $\beta_2=-0.5 \sin \left( \frac{2 \pi r}{r_t}\right)$ . Left panels correspond to $n=5+1$   coefficients $a_i$ with no penalty.    Top left panel:  synthetic data, true and highest likelihood  fit  of $\sigma_{los}^2$.  Bottom left panel: corresponding $\sigma_{rr}^2$ fit and reference values.   
 Yellow shaded regions correspond to the $1\sigma$ uncertainty interval of the unknown coefficients $a_i$ only, i.e. we do not account for the uncertainty of the mass defining parameters $(w_0,\rho_0,rc)$. 
Right panels correspond to the same mass model and anisotropy, however now we consider $n=7+1$ coefficients $a_i$ and the full posterior distribution with penalty $p(W|\lambda)$ and hyperprior $p(\lambda|\alpha,\beta)$ (Equations \ref{DLI_BSplines_penalty_pdf} and \ref{DLI_BSplines_IGprior}). 
}
\label{DLI_BSplines_FigJeans_ΟΜ_fit}
\end{figure*}

\begin{table}
\caption{Bayesian evidence for system with sinusoidal anisotropy $\beta(r)=-\sin(2 \pi r / r_t)$. }
\label{DLI_BSplines_Example3_evidence}
\centering
\begin{tabular}{@{} c | c | c | c }
 \hline 
 Penalty &NDim & Number of coefficients $n$  & $\ln Z \pm \delta \ln Z$ 
 \\ \hline \hline
No & 9  & $n=5+1$     & $-51.46     \pm  0.20 $  \\[6pt]  \hline 
No & 10 & $n=6+1$     & $-55.76\pm 0.20$  \\[6pt]  \hline 
Yes & 12  & $n=7+1$     & $-71.93 \pm 0.22 $  \\[6pt]  \hline 
Yes & 13  & $n=8+1$     & $ -76.25     \pm  0.22 $  \\[6pt]  \hline
 \end{tabular}
\medskip

In this table we list Bayesian evidence values  estimated from MultiNest. For all our fits we used uniform knot distribution and a fixed order $k=5$ of the B-spline basis. 
From left to right: First column describes if we used penalty or not in the corresponding fit. Second column is the total dimensionality of the model. Third refers to the number $n$ of coefficients   $a_i$ of the $\psi\equiv \sigma_{rr}^2(r) = \sum_i B_{i,k}(r)$ representation. Third  column is the value of Bayesian evidence as estimated from MultiNest. The highest value of $\ln Z$ corresponds to the most probable model.  

\end{table}

\subsection{Sinusoidal anisotropy and King mass density}

In this example we consider again a King mass profile, and an anisotropy $\beta$ function of the form:
\begin{equation} \label{DLI_BSplines_beta2}
\beta(r) = -0.5 \sin \left(\frac{2 \pi r}{r_t}\right)
\end{equation}
We do not test the system for stability, or self consistency, i.e. if it is possible to have  a realisation of a stellar distribution with such an anisotropy profile. We are interested to see if the mass estimate of the system is  recovered completely, despite the ``difficult'' anisotropy profile. Namely, the marginalised distributions of the parameters $(w_0,\rho_0,r_c)$ and the mass-to-light ratio $\Upsilon$. We construct our synthetic data in the same way as in the previous examples using $20\%$ error, again allowing the mass-to-light ratio to be a free parameter. The reference values we used for creation of synthetic data are $\{w_0^{ref},\rho_0^{ref},r_c^{ref}, \Upsilon^{ref} \}=\{5,200,5,10\}$.

In Table \ref{DLI_BSplines_Example3_evidence} we list the Bayesian evidence for various fits of our model to synthetic data, with and without penalty. As in the previous example,  the values of  $\alpha,\beta$ in the hyperprior $p(\lambda|\alpha,\beta)$ are the ones that correspond to the estimates from the King Isotropic model (section \ref{DLI_BSplines_ML_detail}). Bayesian inference favours the models with $n=5+1$ for the case with no penalty, and $n=7+1$ for the case with penalty. 

In Fig. \ref{DLI_BSplines_FigJeans_ΟΜ_fit} we plot the line-of-sight velocity dispersion and the  second order radial velocity moment of these models; left panels correspond to fits without penalty, while right panels to  fits with penalty. Yellow shaded region corresponds to the $1\sigma$ uncertainty intervals of the coefficients $a_i$ only, i.e. keeping the mass parameters $(w_0,\rho_0,r_c)$ fixed at the highest likelihood values. Taking into account the variance of the mass model, the true uncertainty would be greater. 
For the case  without penalty, we see that $\sigma_{los}^2$ follows the shape of the true curve, however around $r\sim 10$ deviates  from the true value. Note  that if we included the variance of the mass model as well, the true curve will be overlapped by the uncertainty region of the model. 
The corresponding $\sigma_{rr}^2$ highest likelihood fit fails to follow the slope of the true curve close to the origin. However the true curve lies within the $1\sigma$ uncertainty, except very close to the origin. 
 This is due to the small number of data points and larger errors in the synthetic data. 
 With more data points or smaller errors, as we shall see, we get an excellent fit. 
This discrepancy is also due to the small number of coefficients, $a_i$, we used and the uniform knot distribution, which is far from optimum. We need more control points for the B-spline representation in regions of increased curvature. This is something that the uniform knot distribution fails to encapsulate with only $n=5+1$ coefficients.  
 The case where we  consider  penalty gives a much better $\sigma_{los}^2$  reconstruction. The small deviation in the origin is within the $1\sigma$ uncertainty of the $a_i$ coefficients. The corresponding $\sigma_{rr}^2$ plot  again deviates close to $r=0$. This is due to the bad quality of the data and the bad choice of knot distribution. However the majority of the true curve is encompassed in the $1\sigma$ uncertainty. We discuss a method for overcoming the  problem of the poor fit close to the origin in Section \ref{DLI_BSplines_Discussion}.

\begin{figure}
\centering
%height=0.3\textwidth, width=0.5\textwidth
%\showthe\columnwidth 
\includegraphics[width=\columnwidth]{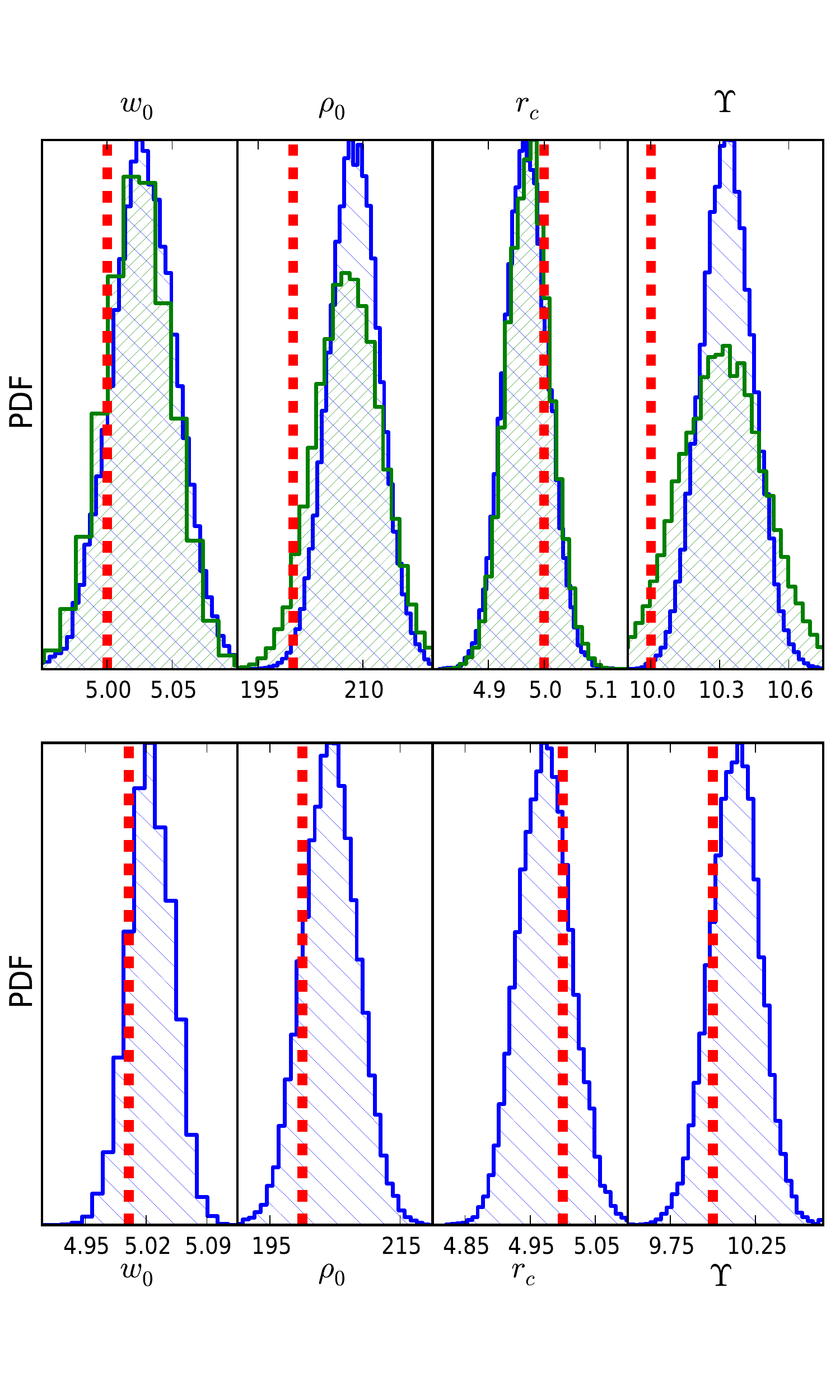}
 \caption{
 Histograms for the system with sinusoidal anisotropy $\beta_2 = - \sin \left(\dfrac{2 \pi }{r_t} r \right) $ and King mass density. 
The red dashed lines correspond to the reference values $(w_0^{ref},\rho_0^{ref},r_c^{ref},\Upsilon^{ref})=(5,200,5,10)$ from which synthetic data were created. Top panel: marginalised distributions of the mass model  parameters $(w_0,\rho_0,r_c,\Upsilon)$ for $n=5+1$ (blue hatched histogram) and  $n=6+1$ (green hatched histogram) coefficients $a_i$ model, with no penalty.
The case with $n=6+1$ has greater variance in all variables, and encapsulates better the reference value of the mass-to-light ratio. 
 Bottom panel:  
 The same mass model parameters, but now for the case of $n=7+1$ coefficients $a_i$ with  smoothing penalty.  In all cases the mass content of the cluster is fully reconstructed, although in the $n=5+1$ case with no penalty, for $\Upsilon$ this is in the total uncertainty of the parameter. }
\label{DLI_BSplines_FigJeans_sin_params}
\end{figure}

\begin{figure}
\centering
%height=0.3\textwidth, width=0.5\textwidth
%\showthe\columnwidth 
\includegraphics[width=\columnwidth]{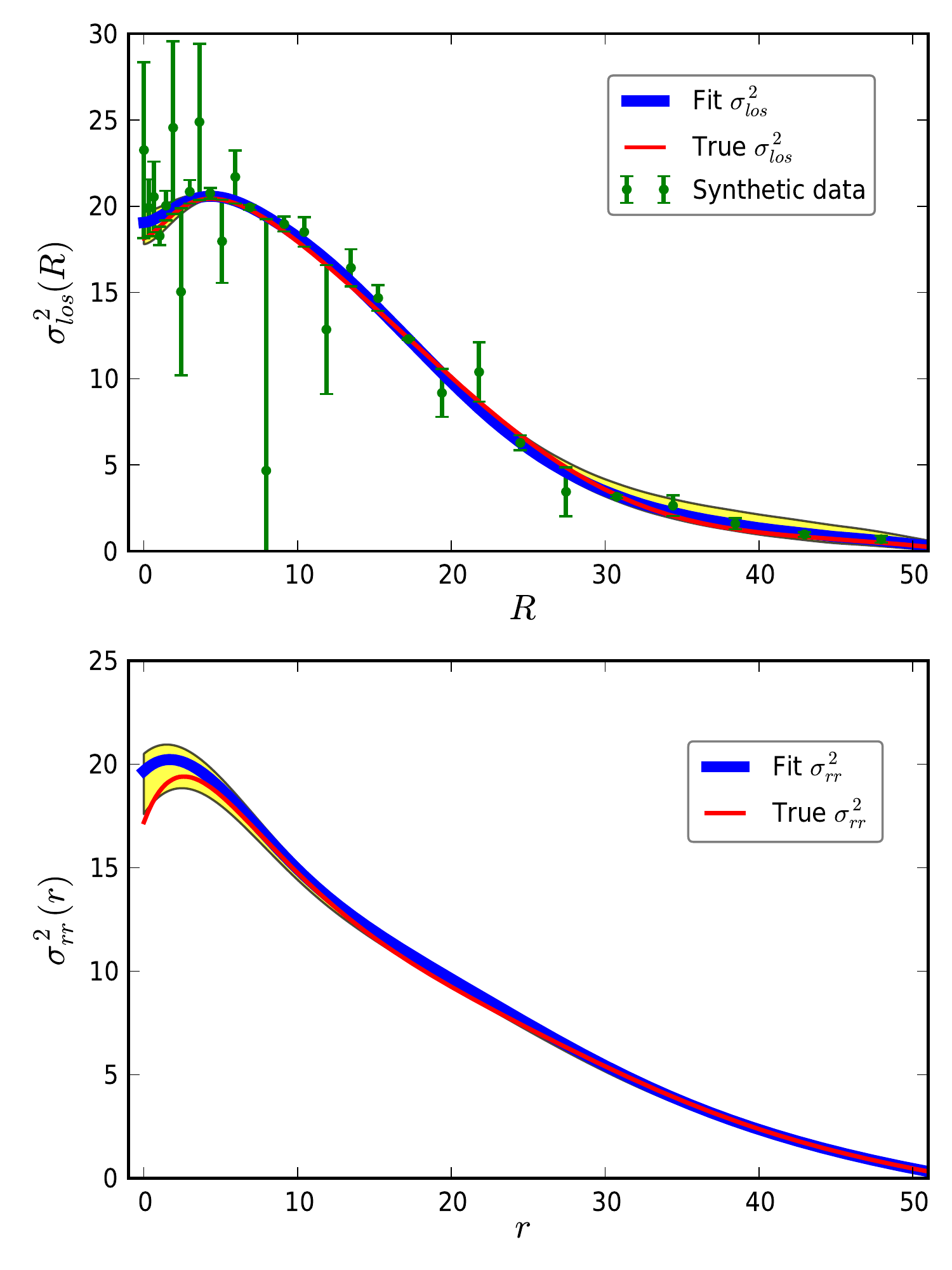}
 \caption{Line-of-sight velocity dispersion synthetic data, true and fit  $\sigma_{los}^2$ (top panel) and second order radial velocity moment $\sigma_{rr}^2$ (bottom panel) fit for the case where synthetic data were created from a sinusoidal profile $\beta(r) = - \sin \left(\dfrac{2 \pi }{r_t} r \right) $. This example  uses 29 data points with $20\%$ random error. As usual the yellow shaded region corresponds to $1\sigma$ uncertainty of the coefficients $a_i$ only, that is, keeping the mass model fixed to the highest likelihood values.} 
\label{DLI_BSplines_FigJeans_sin3_30}
\end{figure}

 In Fig. \ref{DLI_BSplines_FigJeans_sin_params} we plot the histograms of the corresponding mass model parameters, as well as the mass-to-light ratio $\Upsilon$; vertical dashed lines correspond to the reference values from which synthetic data were created, while the top panel corresponds to fit without penalty. The bottom panel to fit with penalty. Observe that in all cases the mass content of the system is reconstructed within the uncertainty of the parameters. In the top panel we plot the marginalized distributions of two models, namely $n=5+1$ (blue in online version), and $n=6+1$ (green in online version). The model with $n=5+1$ coefficients $a_i$ contains the reference values in the outer limits of the distributions for the case of $\rho_0$ and $\Upsilon$. This is due to the reduced flexibility of the model that introduces bias in the estimates. The small number of coefficients with a uniform knot distribution in combination with  the bad quality of our data  fail to describe accurately the region close to $r=0$. The model with $n=6+1$ that is more flexible recovers  the reference values with a smaller uncertainty interval. Note that Bayesian inference  penalizes any complexity resulting from higher number of dimensions. However the difference in $\ln Z$ between competing models with no penalty is not decisive according to Jeffreys Table (see Paper I for further details). That is, the model with $n=6+1$ is acceptable, although less favored.   
 The model with penalty again gives a better fit and encapsulates the reference values in smaller uncertainty.

In order to verify that indeed the deviation from the true curve close to the origin is due to the data set, we recalculated our parameters with penalty using $n=7+1$ coefficients $a_i$ and  29 data points, again with $20\%$ error on the reference values. In Fig. 
\ref{DLI_BSplines_FigJeans_sin3_30} we plot the  $\sigma_{los}^2$ (top panel) and corresponding $\sigma_{rr}^2$ (bottom panel) highest likelihood fits. Indeed in this case the kinematic profile reconstruction is excellent, as we expected. We note that in order to have the most accurate  mass model parameter estimates, we should have recalculated  $(\alpha,\beta)$.  Then  the penalty distribution  would be fine tuned for the larger data set. We should also perform Bayesian model selection again. However this goes beyond our needs for this simple example: we only want to demonstrate that with greater number of data, we have a better fit to the kinematic profile and our algorithm can account for the increased curvature of $\sigma_{rr}^2$ close to the origin.

\subsection{Osipkov-Merritt anisotropy and Cold Dark Matter}

In this example we consider the case of  a combined model that has both a stellar population and a DM component. 
The  stellar component  is  described by a King profile, and the DM  with an NFW profile.  
We create a sample of 14 data points with $10\%$ error, in the same way as in the previous examples.  
The reference values for the creation of synthetic data are $\{w_0^{ref},\rho_{0\star}^{ref},r_c^{ref},\rho^{ref}_{0\bullet}, r^{ref}_s, Q \}=\{5,200,5,20,50,0.5\}$ and the functional form of the anisotropy $\beta(r)$ profile is:
\begin{equation}
\beta(r) = Q 
\frac{r^2}{r_p^2+r^2}
\end{equation} 
 where    $r_p=0.5 (r_c^{ref}+r_s^{ref})$. 
 For the brightness profile we make the approximation that, for each solar mass  $M_{\odot}$ of the tracer mass density corresponds a solar luminosity $L_{\odot}$, i.e. $\Sigma_{\star}(R) \approx J(R)$.   As usual, we consider an adaptive uniform knot distribution, and the order of the B-spline approximation is again $k=5$. 
The coefficients, $\alpha, \beta$, of the hyperprior, $p(\lambda|\alpha,\beta)$,  
were estimated for optimum smoothing from synthetic data based on the Osipkov-Merritt anisotropy model as well as the isotropic King model  (section \ref{DLI_BSplines_ML_detail}). That is, we combine information on smoothness from two distinct models with different anisotropy profiles. We do so in order to allow for more information to be encoded to our model.

For this example we try several models with various numbers of B-spline coefficients $a_i$. We perform model selection using BIC since MultiNest converges too slowly due to the higher dimensionality of parameter space.  The results of the Bayesian inference can be seen in Table \ref{DLI_BSplines_Example4_14_evidence}. When we do not consider a smoothing penalty the most probable model is the one with $n=5+1$ coefficients, $a_i$. The best fitting model with penalty has $n=8+1$ coefficients,  $a_i$. 

\begin{figure*}
\centering
%height=0.3\textwidth, width=0.5\textwidth
%\showthe\columnwidth 
\includegraphics[width=\textwidth]{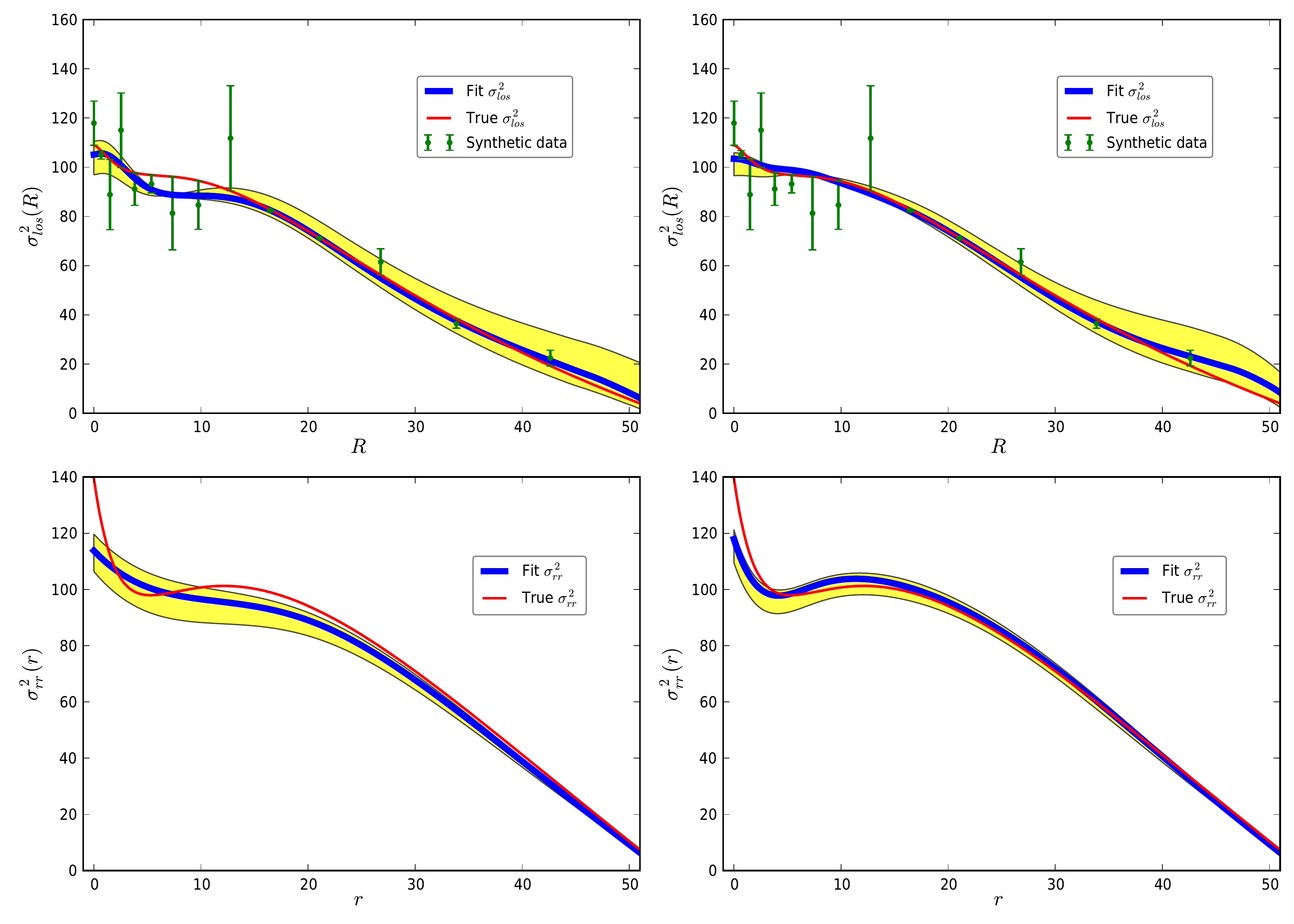}
 \caption{
Fits for the cases of  King tracer mass model with Dark Matter and anisotropy parameter $\beta(r)=0.5  \left( \frac{r^2}{r_p^2+r^2}\right)$ where $r_p = (r_c^{ref}+r_s^{ref})/2$. For this example we used 14 synthetic data points with $10\%$ error. Left panels correspond to $n=5+1$   coefficients $a_i$ with no penalty.    Top left panel:  synthetic data, true and highest likelihood  fit  of $\sigma_{los}^2$.  Bottom left panel: corresponding $\sigma_{rr}^2$ fit and true reference profile.   
 Yellow shaded regions correspond to the $1\sigma$ uncertainty interval of the unknown coefficients $a_i$ only, i.e. we do not account for the uncertainty of the mass defining parameters $(w_0,\rho_{0\star},r_c,\rho_{0\bullet},r_c)$. 
Right panels correspond to the same mass model and anisotropy, however now we consider $n=8+1$ coefficients $a_i$ and the full posterior distribution with penalty $p(W|\lambda)$ and hyperprior $p(\lambda|\alpha,\beta)$ (Equations \ref{DLI_BSplines_penalty_pdf} and \ref{DLI_BSplines_IGprior}). 
}
\label{DLI_BSplines_FigJeansKingDM_fit}
\end{figure*}

We plot the highest likelihood fits in Fig. \ref{DLI_BSplines_FigJeansKingDM_fit}; the left panels correspond to $n=5+1$ coefficients and no penalty. The top left panel is the line-of-sight velocity dispersion data, the fit and the true profile;    bottom left panel is the corresponding $\sigma_{rr}^2$ fit, while  
right panels correspond to profiles with $n=8+1$ coefficients $a_i$ and penalty. The top right panel is the line-of-sight  velocity dispersion data, true profile and highest likelihood fit, while bottom right panel is the corresponding $\sigma_{rr}^2$ fit. 
In all figures, the yellow shaded region corresponds to the $1\sigma$ uncertainty interval of the coefficients $a_i$ only, i.e. keeping the mass models parameters fixed to the highest likelihood value.
The case with  no penalty has more flexibility and tends to follow the data in the region $R\sim 10$. It also fails to capture the increased curvature in the region $R \sim 5$,  thus deviating from the reference. The model with penalty, again gives  a better fit. It has a small deviation close to the origin $R \sim 0-5$. This deviation is due to the small number of data points. 

\begin{table}
\caption{Bayesian Information Criterion (BIC) for Osipkov-Merritt anisotropy and compound system with Stellar and DM components, for the case of 14 data points.}
\label{DLI_BSplines_Example4_14_evidence}
\centering
\begin{tabular}{@{} c|  c | c | c }
 \hline 
Penalty & NDim & Number of coefficients $n$  &BIC  
 \\ \hline \hline
No & 10  & $n=5+1$     & $183.457$  \\[6pt]  \hline 
No & 11  & $n=6+1$     & $185.568$  \\[6pt]  \hline 
 Yes & 13 &  $n=7+1$   & $333.683$  \\[6pt] 
\hline
 Yes & 14 &  $n=8+1$   & $332.951$  \\[6pt] 
\hline
Yes &  15  &$n=9+1$   &  $ 350.699$ 
 \\[6pt]  \hline
Yes & 16  &  $n=10+1$   & $ 355.776 $  \\[6pt]
\hline 
\end{tabular}
\medskip\\
The order of the B-spline representation of $\psi\equiv \sigma_{rr}^2(r)$ was held fixed at $k=5$. 
From left to right: first column describes if we used penalty or not. Second column is the total dimensionality of the model.   Third column is the number of coefficients $a_i$ and the fourth column is the value of BIC. The smallest value of BIC corresponds to the most probable model.  
\end{table}

\begin{figure}
\centering
%height=0.3\textwidth, width=0.5\textwidth
%\showthe\columnwidth 
\includegraphics[width=\columnwidth]{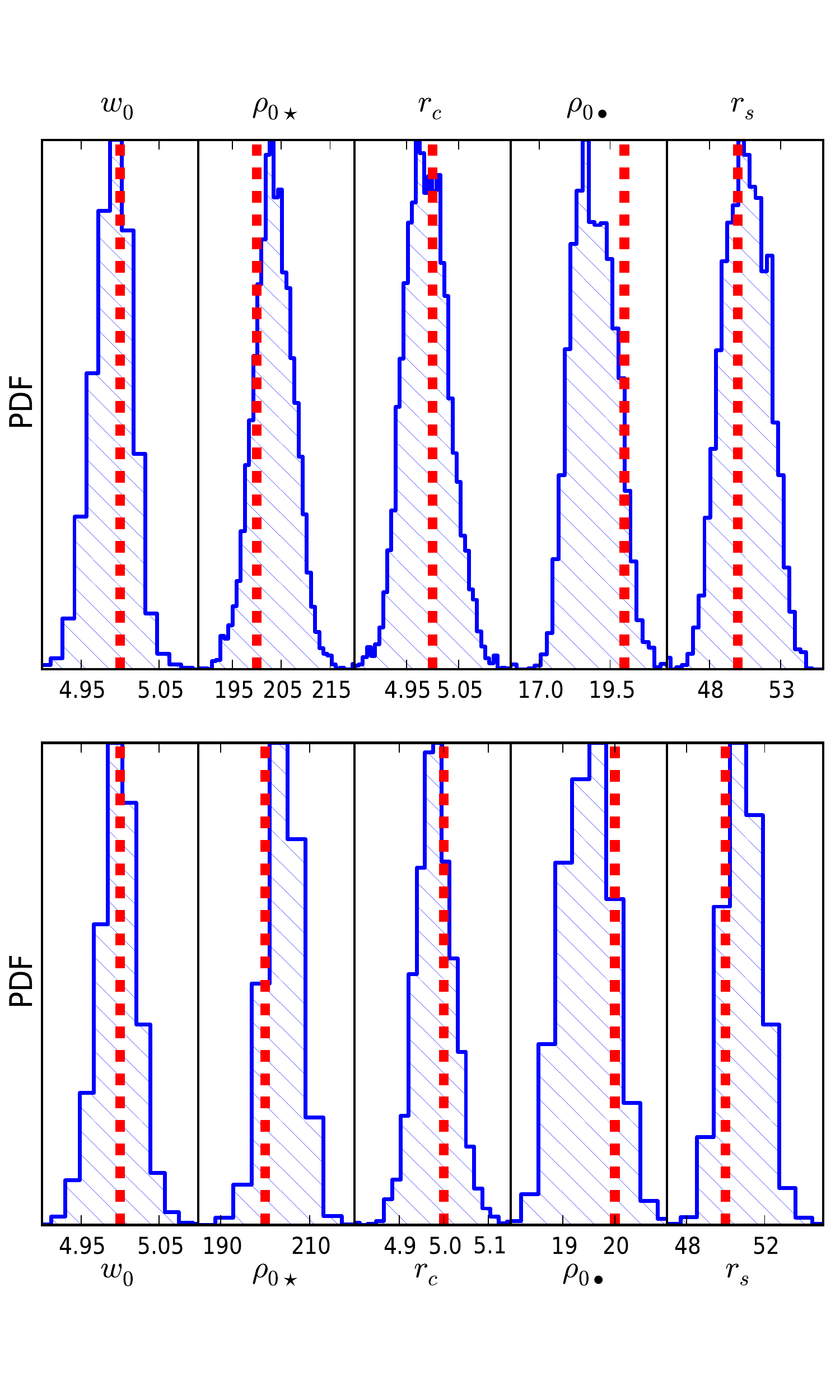}
 \caption{ 
Histograms of composite system with Osipkov - Merritt anisotropy, stellar King profile and NFW profile for DM. In this example we used  14 data points. Top panel has mass models marginalized distributions and corresponds to $n=5+1$ coefficients $a_i$ with no penalty.   Bottom panel: The same with $n=8+1$ coefficients $a_i$ and smoothing penalty distribution. 
}
\label{DLI_BSplines_FigJeansKingDM_param}
\end{figure}

In Fig.  \ref{DLI_BSplines_FigJeansKingDM_param} we plot the histograms  of the defining parameters of the mass models. Vertical dashed lines are the reference values from which we created the synthetic data. The top panel corresponds to the case with no penalty, and the bottom panel to the case with penalty. 
In both cases the mass content of the cluster is fully recovered. This holds even for the case with no penalty where we did not have a good reconstruction of the $\sigma_{rr}^2$ profile.

\section{Discussion}
\label{DLI_BSplines_Discussion}

Having presented our method to its full extent there are several issues to be addressed. 
A first comment to be made is on the mass profile. Why did we choose a King mass profile, and did not allow also for a B-spline representation of the mass density $\rho(r)$? We choose to model with a King profile\footnote{Alternatively we could have used a Michie model \citep{1963MNRAS.125..127M} or any model that can give a nice fit to a specific data set.}, since this is in general a good approximation to real stellar systems and above all a simple mathematical construction. The case where the mass density is also allowed to have a free B-spline functional form  requires a different mathematical formulation and presents different technical difficulties.   This is a method under development that we will present in  future work.

As regards to the knot distribution,  our choice of using uniform knots is far from optimum. This is especially evident in our second example where we use a difficult sinusoidal anisotropy profile that has the greatest curvature close to the origin. Our algorithm can be substantially improved by the appropriate choice of knots. A very promising  approach is to use Genetic Algorithms   
\citep{DBLP:journals/cad/YoshimotoHY03} to identify the best knot distribution for a given data set. Then use this distribution in an MCMC scheme according to what we described in order to recover uncertainties of the model defining parameters. 

For models where we also include a smoothing penalty,   we need to emphasise  the importance of the correct  procedure to obtain parameters $(\alpha, \beta)$ of the hyperprior distribution $p(\lambda|\alpha,\beta)$  (Equation \ref{DLI_BSplines_IGprior}). Failure to do so can result in bias in the marginalised distributions of model parameters. This is similar to the bias inserted from the assumption of a specific $\beta(r)$ anisotropy profile.  However, quantitatively this is much smaller due to the flexibility of the B-spline representation of $\sigma_{rr}^2$ even when we consider a smoothing penalty.  
 
 As an example of bias due to wrong $(\alpha, \beta)$ estimates, we will repeat the fit in the last example, where we considered also a DM component. Using the $(\alpha, \beta)$ parameters we obtained from 14 data points for the cases of isotropic and Osipkov-Merritt anisotropy $\beta(r)$,  we fit another model to a set of 29 data points. As explained in section \ref{DLI_BSplines_ML_detail} we should train our model for optimum smoothness with the same number of data points.  Failure to do so affects the relative contribution of  $p(W|\lambda)$ and $p(\lambda|\alpha,\beta)$ to the resulting value of $\mathcal{L}(D|\theta)$. This happens because the  likelihood product, $\mathcal{L}(D_B|\theta) \mathcal{L}(D_K|\theta)$ (see Paper I), for a larger data set contains more terms, each  smaller than unity.
Then the quantities $p(W|\lambda)$ and $p(\lambda|\alpha,\beta)$ must take a different range of values in order not to have a larger  effect on the resulting likelihood. The maximum value and spread of the normalised $p(\lambda|\alpha,\beta)$  is regulated by the values of $(\alpha,\beta)$. Then these define the relative contribution for a given number of data points. Eventually it is the MCMC process that choses the distribution of $\lambda$ values, but a wrong assumption on this hyperprior can introduce errors in the analysis. 

\begin{figure*}
\centering
%height=0.3\textwidth, width=0.5\textwidth
%\showthe\columnwidth 
\includegraphics[width=\textwidth]{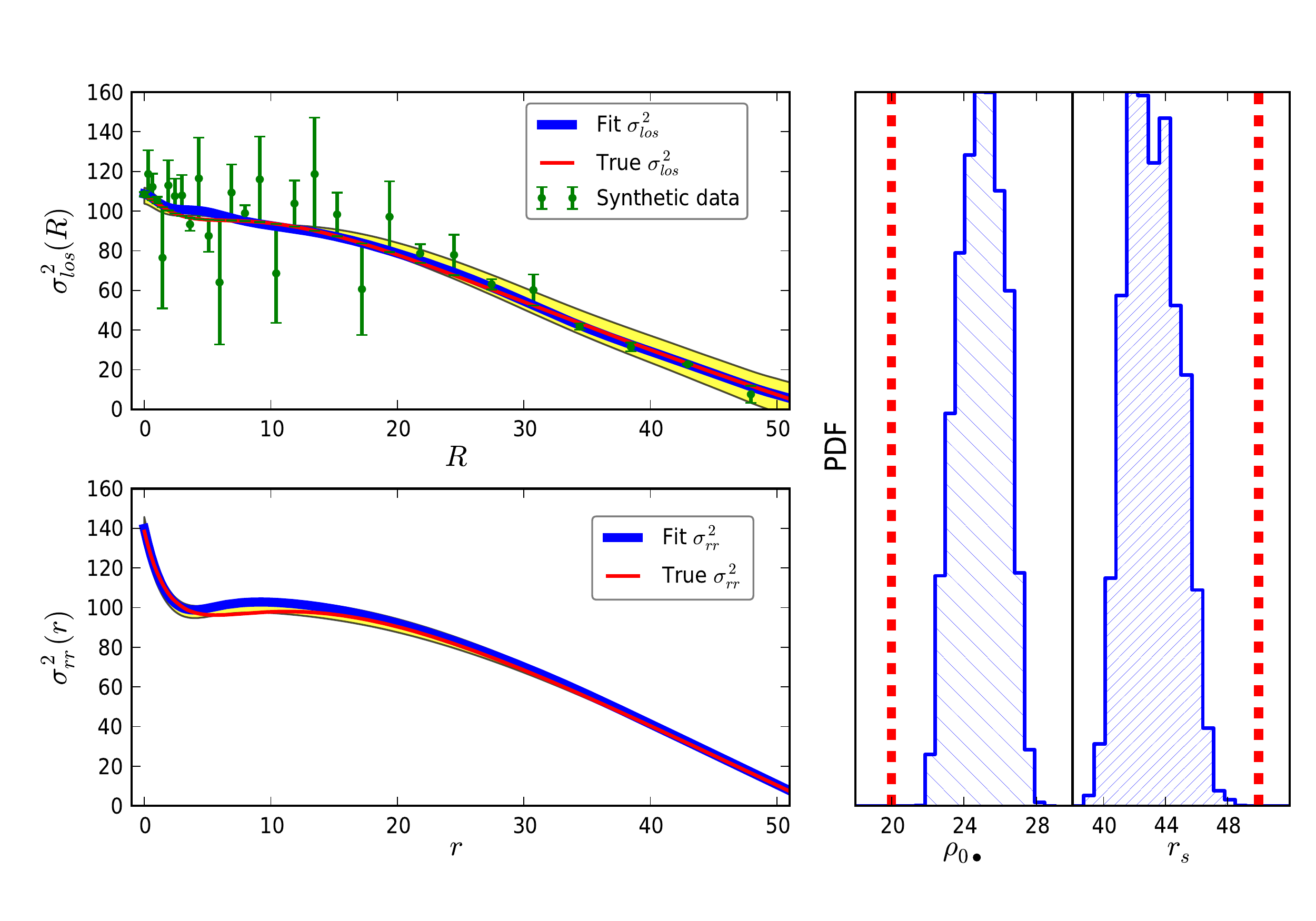}
 \caption{ 
 Fit for the case of King and DM component with Osipkov-Merritt anisotropy, and 29 data points. We choose the highest likelihood fit with $n=10+1$ coefficients $a_i$. The top left panel is the line-of-sight velocity dispersion   $\sigma_{los}^2$ reference profile, fit and synthetic data points. Bottom left panel is the corresponding $\sigma_{rr}^2$ reference profile and fit. Yellow shaded region corresponds to $1\sigma$ confidence interval of the coefficients $a_i$ only, i.e. keeping the mass models fixed to the highest likelihood values.  In the right panel we plot the marginalised distributions of the $(\rho_{0\bullet}, r_s)$ parameters of the  NFW profile. The vertical red lines correspond to the reference values $(\rho_{0\bullet}^{ref}, r_s^{ref})$ from which synthetic data were created. Observe that using  the wrong penalty  parameters $\alpha, \beta$ of $p(\lambda|\alpha,\beta)$ induces bias, even in the case where we use a large number of coefficients $a_i$. 
}
\label{DLI_BSplines_FigJeansKingDM_fit29wpnlty_withbias}
\end{figure*}

In Table \ref{DLI_BSplines_Example4_bias_evidence} we give the BIC values for a set of models with a different number $n$ of coefficients $a_i$. The most probable model results for the case where $n=10+1$. In Fig. \ref{DLI_BSplines_FigJeansKingDM_fit29wpnlty_withbias} we plot the fit of line-of-sight velocity dispersion $\sigma_{los}^2$ (top left panel), the corresponding radial velocity dispersion $\sigma_{rr}^2$ (bottom left panel) and the marginalised distributions of two of the defining parameters of the mass model, namely the DM core density $\rho_{0\bullet}$ and $r_s$ of the NFW profile. The vertical red lines correspond to the reference values from which we created the synthetic data. Observe that despite the fact that we have an excellent fit for the $\sigma_{los}^2$ and $\sigma_{rr}^2$ values, there exists a small bias in the $(\rho_{0\bullet}, r_s)$ estimates. In this example the stellar model parameters were accurately recovered, however we do not plot them here since we only wish to demonstrate the effect of incorrect penalty hyperprior with parameters that have biased marginalised distributions.

As a methodology for obtaining accurate results we propose to always fit with and without penalty to real stellar systems, and observe if there are discrepancies between resulting marginalised distributions of mass model parameters. 
The penalty-free fit will always reconstruct the mass content of a stellar system however it may give unphysical variational $\sigma_{rr}^2$ or $\sigma_{los}^2$ profiles. Then the corresponding penalised fit should result  in marginalised distributions over approximately the same range and with approximately the same variance.

Finally we note that Bayesian inference for model selection heavily penalises the complexity of the models that have a greater number of coefficients and a penalty distribution $p(W|\lambda) p(\lambda|\alpha,\beta)$. Specifically in all examples,  Bayesian inference predicts decisively that models with  penalty should  not be considered. However, in all cases we demonstrated that fits with smoothing penalty give better results. This is a drawback that results from 
our inadequacy to include in the likelihood analysis (in the case where there is no penalty) the information of smoothness that the various functions of a physical system must satisfy. That is, Bayesian inference does not have any information as to  why we impose a smoothing penalty distribution and penalises it.  
This is also directly related to our prior belief of the $\lambda$ penalty parameter (Equations \ref{DLI_BSplines_penalty_pdf} and \ref{DLI_BSplines_IGprior}).

\section{Conclusions}
\label{DLI_BSplines_Conclusions}

\begin{table}
\caption{BIC for Osipkov-Merritt anisotropy and compound system with stellar and  DM components, for the case of 29 data points. }
\label{DLI_BSplines_Example4_bias_evidence}
\centering
\begin{tabular}{@{} c|  c | c | c }
 \hline 
Penalty & NDim & Number of coefficients $n$  &BIC  
 \\ \hline \hline
 \begin{comment}
No & 10  & $n=5+1$     & $439.502$  \\[6pt]  \hline 
No & 11  & $n=6+1$     & $440.207$  \\[6pt]  \hline 
 No & 12 &  $n=7+1$   & $450.738$  \\[6pt] 
\hline\hline
\end{comment}
 Yes & 14 &  $n=8+1$   & $602.067$  \\[6pt] 
\hline
Yes &  15  &$n=9+1$   &  $ 601.618$ 
 \\[6pt]  \hline
Yes & 16  &  $n=10+1$   & $ 600.283$  \\[6pt]\hline
Yes & 17  &  $n=11+1$   & $ 601.194$  \\[6pt]\hline
Yes & 18  &  $n=12+1$   & $603.704$  \\[6pt]
\hline
\end{tabular}
\medskip\\
The order of the B-spline representation of $\psi\equiv \sigma_{rr}^2(r)$ was held fixed at $k=5$. 
From left to right: first column describes if we used penalty or not. Second column is the total dimensionality of the model.   Third column is the number of coefficients $a_i$ and the fourth column is the value of BIC. The smallest value of BIC corresponds to the most probable model.  
\end{table}

In this work we validate and expand the method developed in Paper I. Namely we address the issue of the mass-anisotropy degeneracy of the spherically symmetric Jeans equation. We  present an algorithm that combines smoothing B-splines with dynamical equations of physical systems and reconstructs accurately the kinematic profile and the mass content of a stellar system. This is for a constant or variable mass-to-light ratio $\Upsilon$.

Based on the assumption that a realistic physical system must have similar behaviour 
to ideal theoretical models 
in terms of smoothness of the  $\sigma_{los}^2$ function,  we present a method for estimation of the optimum smoothing penalty  for a statistical model fitting of the B-spline representation of $\sigma_{rr}^2$. Furthermore we demonstrate with an example that incorrect smoothing penalty estimation can lead to biases in the defining parameters of the mass models. 

We present three examples of kinematic profile reconstruction from brightness and line-of-sight velocity dispersion observables. These are based on synthetic data  that consist of 14 data points, each with a Gaussian random error  on the reference value.   The first two examples have a constant mass-to-light ratio $\Upsilon$, while the third example consists of a composite structure with a stellar and a DM component. In all cases we reconstruct completely the mass content of the system and the kinematic $\sigma_{rr}^2$ and $\sigma_{los}^2$ profile.  

Thus we removed the mass-anisotropy degeneracy to the level that for an assumed free functional form of the potential and mass density pair $(\Phi(r), \rho(r))$, and a given  data set of brightness, $J$, and line-of-sight velocity dispersion observables, $\sigma_{los}^2$,  we reconstruct a unique kinematic profile  $(\sigma_{rr}^2,\sigma_{tt}^2)$, within the statistical uncertainties. 
In Paper I we argued that if one knows the complete functional form of the line-of-sight velocity dispersion $\sigma_{los}^2$ and the mass density\footnote{This is valid also for cases where we have a composite system that consists of a stellar, $\rho_{\star}(r)$, and a dark matter component, $\rho_{\bullet}(r)$, i.e. $\rho(r)=\rho_{\star}(r)+\rho_{\bullet}(r)$.} $\rho(r)$, then it is in principle possible to find a unique decomposition of $\sigma_{los}^2$ to $\sigma_{rr}^2$ and $\sigma_{tt}^2$. For the case of discrete data we can speak only for a unique family of solutions within some statistical uncertainty. 
This uniqueness is identified through the unimodal marginalised distributions of the model parameters from the MCMC scheme.

In general the quality of our results depends on the quality of the data. That is, on the total number of binned brightness and $\sigma_{los}^2$ observables and their corresponding errors. The method was demonstrated to give excellent results for as few as 14 binned data points and up to $20\%$ error in the reference value from which we created the synthetic data. \\

\section*{Acknowledgments}
F. I. Diakogiannis acknowledges the University of Sydney International Scholarship
for the support of his candidature. 
G. F. Lewis acknowledges support from ARC Discovery Project (DP110100678) and Future Fellowship (FT100100268). 
The authors would like to thank Nick Bate as well as the anonymous referee for useful comments and suggestions on the manuscript. 

%
%\nocite{*}
%\bibliographystyle{plain}
%\bibliography{references}
%\nocite{*} 
%\begin{thebibliography}{99}
\bibliographystyle{mn2e} %-> the bibliography style for MNRAS
\bibliography{DLI_BSplines_II.bib} %-> YourListOfArticle is the compiled version of %YourListOfArticle.bib 
%\end{thebibliography}

%\appendix

\bsp

\label{lastpage}

\end{document}